\documentclass{emulateapj}

\newcommand{\Msun}{M$_{\odot}$}

\def\gsim{\mathrel{\rlap{\lower 4pt \hbox{\hskip 1pt $\sim$}}\raise 1pt
\hbox {$>$}}}
\def\lsim{\mathrel{\rlap{\lower 4pt \hbox{\hskip 1pt $\sim$}}\raise 1pt
\hbox {$<$}}}

\begin{document}

\title{Nucleosynthesis in Two-dimensional Delayed Detonation Models \\ 
of Type Ia supernova Explosions}

\author{
K.~Maeda\altaffilmark{1,2}, 
F.K.~R\"opke\altaffilmark{2}, 
M.~Fink\altaffilmark{2}, 
W.~Hillebrandt\altaffilmark{2}, \\
C.~Travaglio\altaffilmark{2,3}, 
F.-K.~Thielemann\altaffilmark{4,5}}

\altaffiltext{1}{Institute for the Physics and Mathematics of the 
Universe (IPMU), University of Tokyo, 
5-1-5 Kashiwanoha, Kashiwa, Chiba 277-8583, Japan; 
keiichi.maeda@ipmu.jp.}
\altaffiltext{2}{Max-Planck-Institut f\"ur Astrophysik, 
Karl-Schwarzschild-Stra{\ss}e 1, 85741 Garching, Germany.}
\altaffiltext{3}{INAF-Osservatorio Astronomico di Torino, 
Strada dell'Osservatorio 20, I-10025 Pino Torinese, Torino, Italy}
\altaffiltext{4}{Department Physik, Universit\"at Basel, 
CH-4056 Basel, Switzerland}
\altaffiltext{5}{GSI Helmholtz Center, 
Planckstr. 1, 64291 Darmstadt, Germany} 

\begin{abstract}
For the explosion mechanism of Type Ia supernovae (SNe Ia), different
scenarios have been suggested. In these, the propagation of
the burning front through the exploding white dwarf star proceeds in
different modes, and consequently imprints of the explosion model on
the nucleosynthetic yields can be expected. The nucleosynthetic
characteristics of various explosion mechanisms is explored based on
three two-dimensional explosion simulations representing extreme
cases: a pure turbulent deflagration, a delayed detonation following an 
approximately spherical ignition of the initial deflagration, and a delayed
detonation arising from a highly asymmetric deflagration ignition. 
Apart from this
initial condition, the deflagration stage is treated in a 
parameter-free approach.  
The detonation is initiated when the turbulent burning enters the distributed 
burning regime. This occurs at densities around $10^{7}$ g cm$^{-3}$ -- 
relatively low as compared to 
existing nucleosynthesis studies for one-dimensional spherically 
symmetric models. The burning in these multidimensional models
is different from that in one-dimensional simulations as
the detonation wave propagates both into unburned material in 
the high density region near the 
center of a white dwarf and into the low density region near the surface. 
Thus, the resulting yield is a mixture of different explosive 
burning products, from carbon-burning products at low densities to 
complete silicon-burning products at the highest densities, as well as 
electron-capture products synthesized at the deflagration stage.
Detailed calculations of the nucleosynthesis in all three models are
presented.  
In contrast to the deflagration model, the delayed detonations produce
a characteristic layered structure and the yields largely satisfy
constraints from Galactic chemical evolution. In the asymmetric
delayed detonation model, the region filled with electron capture species
(e.g., $^{58}$Ni, $^{54}$Fe) 
is within a shell, showing a large off-set, 
above the bulk of $^{56}$Ni distribution, 
while species produced by the detonation are distributed more spherically. 
\end{abstract}


\section{INTRODUCTION}

There is a consensus that Type Ia supernovae (SNe Ia) 
are the outcome of a thermonuclear explosion of a 
carbon-oxygen white dwarf (WD) 
(e.g., Wheeler et al.\ 1995; Nomoto et al.\ 1997; 
Branch\ 1998). 
For the progenitor, the Chandrasekhar-mass ($M_{\rm Ch}$) 
WD model has 
been favored 
for a majority of SNe Ia (e.g., H\"oflich \& Khokhlov\ 1996; 
Nugent et al.\ 1997; Fink et al.\ 2007; Mazzali et al.\ 2007). 

The mass of the WD could reach $M_{\rm Ch}$ 
by several evolutionary paths, either by 
a mass-transfer from a binary giant/main-sequence 
companion (single degenerate scenario; e.g., 
Whelan \& Iben\ 1973; Nomoto\ 1982) 
or as a result of merging with a binary 
degenerate WD companion (double degenerate scenario; 
e.g., Iben \& Tutukov\ 1984; Webbink\ 1984). 
As the WD has accreted a sufficient amount of material, 
the central density of the WD increases and the heating 
rate by the carbon fusion exceeds the 
cooling rate by neutrino emission. 
The evolution is then followed 
by a simmering phase with convective carbon burning, 
lasting for about a century. At the end of this simmering phase, 
the temperature rises to the point where convection can no longer 
efficiently transport away the energy produced by the carbon burning. 
This is a stage where the burning becomes dynamical, initiating a 
thermonuclear flame that propagates outward and disrupts the WD, 
i.e., a supernova explosion 
(Nomoto et al.\ 1984; Woosley \& Weaver\ 1986). 

Once the thermonuclear flame is ignited, there are two possible modes 
of the propagation: subsonic deflagration and supersonic detonation. 
A prompt ignition of the detonation flame is disfavored 
because the resulting nucleosynthesis yield conflicts 
with Galactic chemical evolution (Arnett\ 1969) 
and fails
to produce the strong intermediate-mass element features observed 
in SNe~Ia. Thus, the explosion should start with 
a subsonic deflagration flame. The deflagration stage may 
last until the end of the explosion (the deflagration model; Nomoto et al.\ 1984), 
while it is also possible that the deflagration flame turns into 
the detonation wave [the delayed detonation model, or the 
deflagration-detonation transition (DDT) model; Khokhlov\ 1991; 
Yamaoka et al.\ 1992; Woosley \& Weaver\ 1994; Iwamoto et al.\ 1999]. 

The supernova explosion phase has been investigated by ``classical'' one-dimensional 
spherically symmetric models; the classical deflagration 
W7 model of Nomoto et al. (1984) has successfully explained the
basic contribution of SNe Ia to Galactic chemical evolution, 
as well as basic features of observed spectra and light curves 
of individual SNe Ia of a normal class (Branch et al.\ 1985). 
Some improvement in these observational 
aspects has been obtained by introducing a delayed detonation 
(e.g., H\"oflich \& Khokhlov\ 1996; Iwamoto et al.\ 1999). 

These models, however, treat the propagation speed 
of the deflagration flame as a parameter. Moreover, 
the deflagration flame is hydrodynamically unstable and 
non-sphericity is thus actually essential (e.g., 
Niemeyer et al.\ 1996). 
Recent investigations of the explosion models have intensively 
addressed these issues (e.g., Reinecke et al.\ 2002; 
Gamezo et al.\ 2003; R\"opke \& Hillebrandt\ 2005; 
Bravo \& Garc\'ia-Senz\ 2006; R\"opke et al.\ 2006b; Schmidt \& 
Niemeyer\ 2006). With high resolution 
multi-dimensional hydrodynamic simulations, coupled with 
an appropriate sub-grid model to capture turbulence effects on unresolved scales, 
recent studies 
provide essentially ``parameter-free'' simulations for the initial 
deflagration stage, 
where only the structure of the pre-supernova WD and 
the distribution of the initial deflagration ignition sparks set up the 
initial conditions. 

The multi-dimensional simulations have been performed with different 
initial conditions to see if SNe Ia in general can be explained 
in a framework of a pure deflagration explosion, as is summarized by 
R\"opke et al. (2007b). Detailed nucleosynthesis 
calculations have been performed for some deflagration models 
(Travaglio et al.\ 2004a, 2005; 
Kozma et al.\ 2005, R\"opke et al.\ 2006a). 
These studies indicate that the pure deflagration explosion can explain 
a part of SNe Ia, up to relatively weak and faint SNe Ia in the normal 
population. 
However, it has also been shown that a subsequent detonation phase
is probably necessary  
to account for typical normal SNe Ia and brighter ones 
(Gamezo et al.\ 2005; R\"opke et al.\ 2007b).  

Delayed detonation models have also been investigated with multi-dimensional simulations 
(Gamezo et al.\ 2005; Plewa\ 2007; 
Golombek \& Niemeyer\ 2005; R\"opke \& Niemeyer\ 2007c; Bravo \& Garc\'ia-Senz\ 2008). 
However, compared to the deflagration models as
summarized above, the investigation of
multi-dimensional delayed detonation models is still at the very initial stage. 
The radiation transfer based on the multi-dimensional models has been 
examined by Kasen et al.\ (2009). 
Detailed nucleosynthesis studies have rarely been done 
(but see Bravo \& Garc\'ia-Senz\ 2008). 
In this paper, we present results from detailed nucleosynthesis 
calculations, based on two-dimensional delayed detonation models and, for comparison,
a pure deflagration model.

The two delayed detonation models presented here can be regarded as
extreme cases -- not necessarily with respect to their $^{56}$Ni
production and brightness, but with respect to symmetries/asymmetries
in the explosion phase. While in one model, the deflagration was
ignited in an approximately 
spherical configuration at the center of the WD, the
other model features an off-center ignition and propagation of the initial
deflagration flame similar to the three-dimensional simulations of 
R\"opke et al.\ (2007d). 
It has been suggested that the convection in the simmering phase may be 
dominated by a dipolar mode, and there is a good possibility that the 
deflagration is initiated in an off-center way 
(e.g., Woosley et al.\ 2004, Kuhlen et al.\ 2006). 
In the pure-deflagration model, 
such an explosion cannot account for normal SNe Ia, because the small 
burning surface area should result in inefficient production of $^{56}$Ni 
(see, e.g., R\"opke et al.\ 2007d). 
However, this can be possibly overcome in the delayed detonation scenario, 
as the detonation can potentially produce a large amount of $^{56}$Ni.  

The paper is organized as follows. In \S 2, we present methods 
and models,
in \S 3, we present our results. Discussion of these results from a view point of 
the chemical evolution is given in
\S 4. 
The paper is closed in \S 5 with concluding remarks. 

\begin{deluxetable*}{ccccccc}
 \tabletypesize{\scriptsize}
 \tablecaption{Mass ($M_{\odot}$) and Energetic ($10^{51}$ erg)
 \label{tab:tab_ene}}
 \tablewidth{0pt}
 \tablehead{
   \colhead{Form}
 & \colhead{W7}
 & \colhead{C-DEF}
 & \colhead{C-DDT}
 & \colhead{O-DDT}
 & \colhead{C-DDT/def.\tablenotemark{a}}
 & \colhead{O-DDT/def.\tablenotemark{a}}
}
\startdata
$M_{\rm burn}$\tablenotemark{b} & 1.21  & 0.61 & 1.01 & 1.25 & 0.522 & 0.393\\
$E_{\rm nuc}$\tablenotemark{c} & 1.78 & 0.91 & 1.46 & 1.80 & 0.767 & 0.522\\
$E_{\rm K}$\tablenotemark{d} & 1.28 & 0.41 & 0.96 & 1.30 & \nodata & \nodata\\ 
$E_{\rm K}$ (hyd)\tablenotemark{e}  & 1.30 & 0.41 & 0.93 & 1.27 & \nodata & \nodata
\enddata
\tablenotetext{a}{At the moment when the first DDT takes place.}
\tablenotetext{b}{The incinerated masses. }
\tablenotetext{c}{The nuclear energy generation derived by the postprocess calculations. }
\tablenotetext{d}{The asymptotic kinetic energy resulting from $E_{\rm nuc}$. The progenitor WD 
binding energy is $E_{\rm bind} = 0.505$. }
\tablenotetext{e}{The asymptotic kinetic energy calculated in the hydrodynamic simulations 
with the simplified treatment of the energy generation. }
\end{deluxetable*}

\section{Methods and Models}

\subsection{Explosion Models}
In this paper, we concentrate on two-dimensional models. 
Although some results might be affected by the imposed symmetry
(e.g., Travaglio et al.\ 2004a; R\"opke et al.\  2007b), we believe that examining 2D models
is a natural step forward. Moreover, it illustrates how 
nucleosynthesis in the different models proceeds, and highlights
differences with respect to one-dimensional models. 

In particular, we focus on three models in this paper as follows: 
\begin{itemize}
\item{\bf C-DEF: }
A globally spherically symmetric pure-deflagration explosion 
in which the deflagration was ignited within the \textit{c3}-shape boundary, 
as was done by Reinecke et al.\ (1999a). 
About $2\times10^{-2}\,M_{\odot}$ were initially incinerated to trigger the deflagration. 
This model is similar to one presented already in Travaglio et al.\ (2004a). 
\item
{\bf C-DDT: }
A delayed detonation model which follows the 2D spherical deflagration C-DEF model. 
A prescription for the DDT is given below. 
\item
{\bf O-DDT: }
A delayed detonation model which follows an extremely off-center deflagration. 
The deflagration is ignited by 29 bubbles distributed within an opening
angle with respect to the $z$-axis of 45 degrees. The outermost 
bubble was placed at a distance of $\sim 180$ km from the center.
About $1\times 10^{-5}\,M_{\odot}$ were initially incinerated to trigger the deflagration. 
The DDT is treated in the same way as in the C-DDT model. 
\end{itemize}

The exploding WD had a central density of  $2.9 \times 10^{9}\,\mathrm{g}\,
\mathrm{cm}^{-3}$. 
In the hydrodynamic explosion simulations, burning was treated in 
a simplified way. Only five species (carbon, oxygen, 
alpha-particles, a representative for iron-peak elements, and a 
representative for intermediate mass elements) were followed. The 
deflagration and detonation fronts were modeled with the level-set 
approach (Reinecke et al.\ 1999b;  Golombek \& Niemeyer\ 2005; 
Roepke \& Niemeyer\ 2007c). After passage of the 
zero-level set representing the combustion waves, the material was 
converted from the carbon/oxygen fuel mixture 
to an approximate nuclear ash composition. At high densities, 
NSE is reached and the ash was modeled as a temperature- and 
density-dependent mixture of iron-peak elements and 
alpha-particles. In the NSE region, electron captures neutronizing the 
ashes were followed. 
In the hydrodynamic simulations, $512 \times 512$ cells are used for 
C-DEF and C-DDT models, and $1024 \times 512$ cells are used for 
O-DDT model. 

In the delayed detonation models, we assumed the
deflagration-to-detonation transition to take place once the flame
enters the distributed burning regime (e.g., R\"opke \& Niemeyer\ 2007c)
at a fuel density of $\le 1 \times 10^{7}\,\mathrm{g}\,
\mathrm{cm}^{-3}$ (hereafter $\rho_{\rm DDT}$). 
We emphasize that in
contrast to  the classical 1D models, the DDT is not exclusively
parameterized by $\rho_{\rm DDT}$, but the requirement on reaching the
distributed burning implies also a turbulence criterion (see R\"opke\ 2007a 
for an evaluation in three-dimensional models).
An estimate for entering the distributed regime
is the equality of the laminar flame width and the Gibson scale -- the
scale at which the turbulent velocity fluctuations equal the laminar
flame speed. When the Gibson scale becomes smaller than the laminar
flame width, turbulence affects (and ultimately destroys) the laminar
flame structure. This is a prerequisite for DDT, however, it is not a
sufficient criterion (e.g. Woosley\ 2007; Woosley et al.\ 2009). The
microphysics of DDT is not yet fully known, and therefore we apply
the necessary condition of entering the distributed regime here
only. We emphasize that the hydrodynamic models are still in a
preliminary stage and used here only in order to demonstrate the
nucleosynthesis associated with them. For more robust predictions of
the $^{56}$Ni production and the implied brightness of the events,
more elaborate hydrodynamical simulations should be used 
(F.K. R\"opke et al.,\ in prep.); the models presented here are understood as a
case study demonstrating typical nucleosynthesis for the different
explosion processes.

Results on some synthetic observables (e.g., light curves) derived for similar
two-dimensional models can be found in Kasen et al. (2009), 
and details on hydrodynamic calculations will be presented elsewhere 
(F.K. R\"opke et al.,\ in prep.). 
Note that the present models cover extreme cases in the sequence 
of ``classical'' delayed detonation models where the DDT takes place before
the deflagration wave reaches the WD surface (e.g., Kasen et al. 2009). 
For the situation in which only a few initial bubbles are distributed 
within a small solid angle with a large off-set, the 
``gravitationally confined detonation'' model (e.g.\ Jordan et al. 2008; 
Meakin et al. 2009) has been suggested as an alternative explosion
mechanism. This would be a more extreme case than our O-DDT model.

\subsection{Nucleosynthesis}
We apply the tracer particle method 
to the calculations of nucleosynthesis. The essence is 
to follow thermal histories of Lagrangian particles, 
which are passively advected in hydrodynamic simulations, 
and then to employ detailed nuclear reaction network 
calculations to each particle separately. 
The method was first applied to core-collapse 
supernovae (Nagataki et al.\ 1997; Maeda et al.\ 2002; Maeda \& Nomoto\ 2003), 
and has become popular in the field thanks to its 
simplicity and its applicability to multi-dimensional problems 
(e.g., Travaglio et al.\ 2004b). Travaglio et al. (2004a) applied 
the method to multi-dimensional, purely deflagration explosion 
models of SNe~Ia. 

In the setup for our hydrodynamic simulations, 
$80^{2}$ tracer particles are distributed uniformly in mass coordinate, 
such that each particle represents 
the same mass of $\sim 2.2 \times 10^{-4} M_{\odot} (=M_{\rm wd}/6400)$. 
The particles are advected passively, following the velocity 
field at each time step of the Eulerian hydrodynamic simulations. 
The thermal history each particle experiences is recorded.
The number of the tracer particles is sufficient
to accurately follow the nucleosynthesis. Seitenzahl et al.\ (in
preparation) find that in two-dimensional SN~Ia simulations with
$80^{2}$ tracer particles all isotopes with abundances higher than
$\sim 10^{-5}$ are reproduced with an accuracy of better than 5\%
(except for $^{20}$Ne). 

The nuclear postprocessing calculations are then performed 
for each particle separately. 
To this end, we recalculate the temperature from the recorded
internal energy,  
rather than directly using the value obtained by the hydrodynamic 
simulations (see Travaglio et al.\ 2004a). 
In deriving the temperature, the electron fraction ($Y_{\rm e}$) is assumed to be 0.5, 
which introduces some errors when the electron captures are very active. 
When $T_{9} = T/10^9$ K $ > 6$, we follow the abundance evolution by applying the 
Nuclear Statistical Equilibrium (NSE) abundance 
(i.e., the abundance specified by $\rho$, $T$, and $Y_{\rm e}$) 
rather than fully solving the reaction network, as the NSE is 
reached in this regime of high density and temperature. 
In order to correctly follow the evolution of $Y_{\rm e}$, weak
interactions are computed along with the NSE abundance.  

\begin{figure*}
\begin{center}
        \begin{minipage}[]{0.15\textwidth}
                \epsscale{1.0}
                \plotone{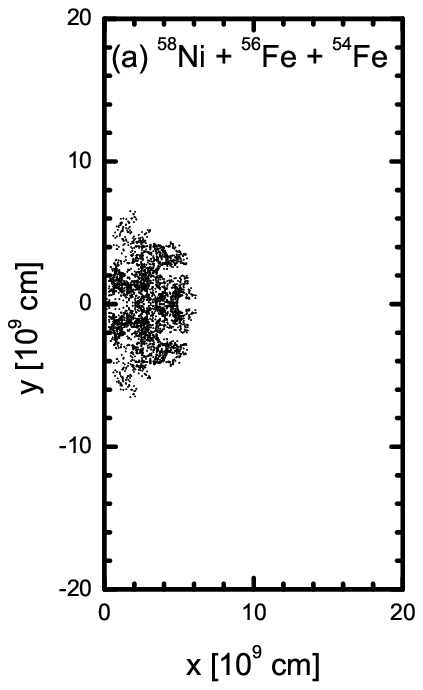}
        \end{minipage}
       \begin{minipage}[]{0.15\textwidth}
                \epsscale{1.0}
                \plotone{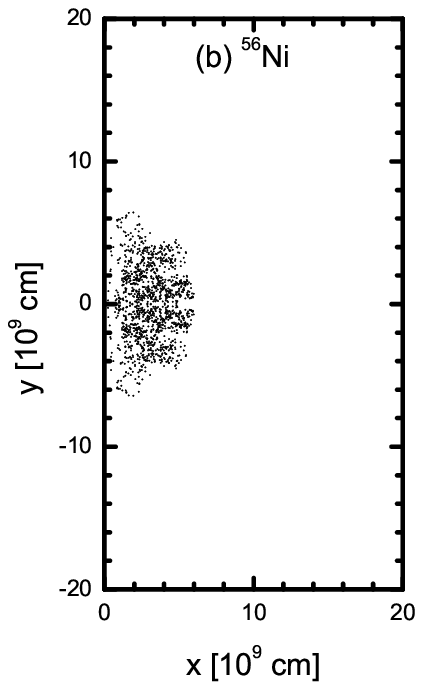}
        \end{minipage}
       \begin{minipage}[]{0.15\textwidth}
                \epsscale{1.0}
                \plotone{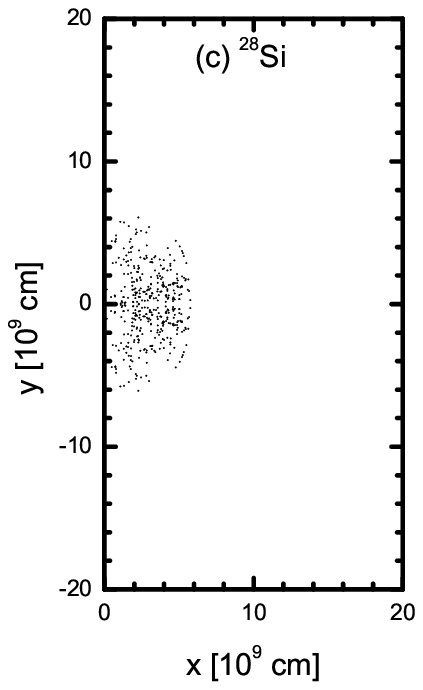}
        \end{minipage}
       \begin{minipage}[]{0.15\textwidth}
                \epsscale{1.0}
                \plotone{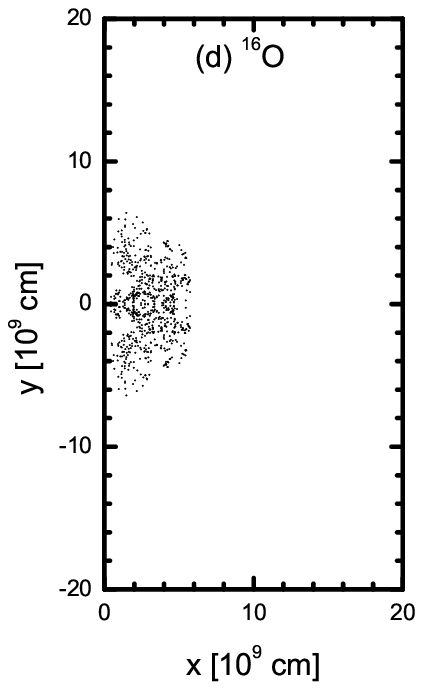}
        \end{minipage}
       \begin{minipage}[]{0.15\textwidth}
                \epsscale{1.0}
                \plotone{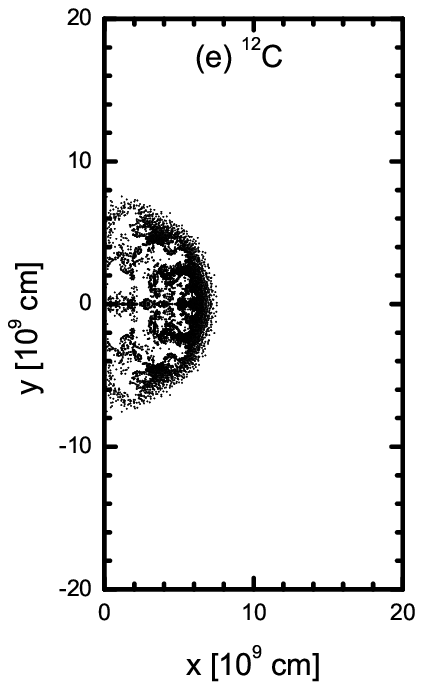}
        \end{minipage}
\end{center}
\caption
{Distribution of characteristic elements/isotopes in the C-DEF model, 
at 10 seconds after the ignition of the deflagration wave. 
(a) Tracer particles in which the total mass fraction of 
stable $^{56}$Fe, $^{58}$Ni, and $^{54}$Fe exceeds 0.1 (electron capture products). 
(b) Tracer particles in which the mass fraction of $^{56}$Ni exceed 0.1, 
excluding particles shown in (a) (complete Si burning). 
(c) Tracer particles in which the mass fraction of $^{28}$Si exceeds 0.1 and 
that of $^{16}$O is below 0.1, excluding particles in (a) and (b) (O burning). 
(d) Tracer particles in which the mass fraction of $^{16}$O exceeds 0.1 and 
that of $^{12}$C is below 0.1, excluding particles in (a) -- (c) (C burning). 
(e) Tracer particles with the mass fraction of $^{12}$C exceeding 0.1, 
excluding particles in (a) -- (d) (no-burning). 
\label{fig:fig1}}
\end{figure*}

\begin{figure*}
\begin{center}
        \begin{minipage}[]{0.15\textwidth}
                \epsscale{1.0}
                \plotone{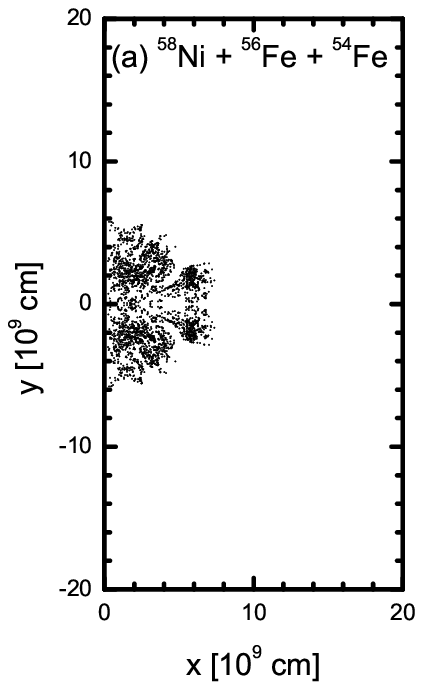}
        \end{minipage}
       \begin{minipage}[]{0.15\textwidth}
                \epsscale{1.0}
                \plotone{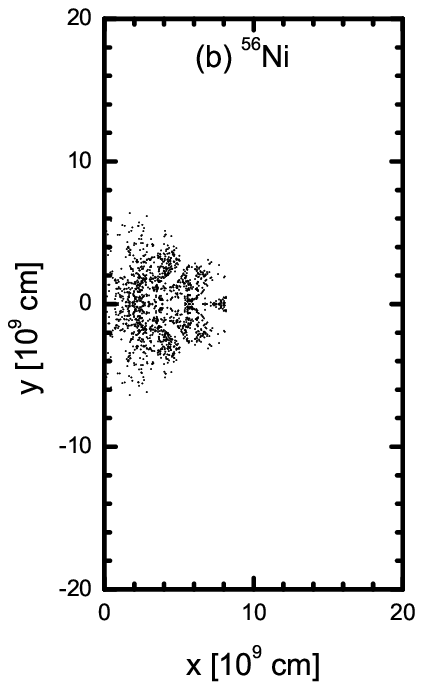}
        \end{minipage}
       \begin{minipage}[]{0.15\textwidth}
                \epsscale{1.0}
                \plotone{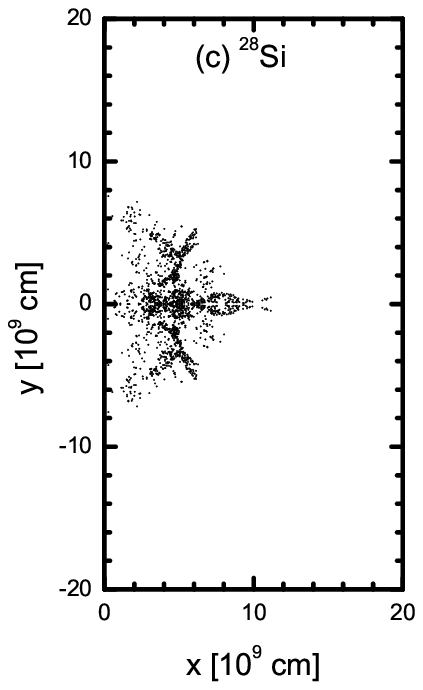}
        \end{minipage}
       \begin{minipage}[]{0.15\textwidth}
                \epsscale{1.0}
                \plotone{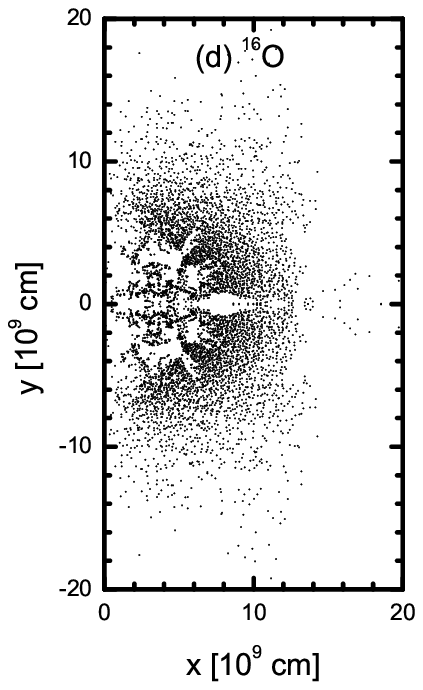}
        \end{minipage}
       \begin{minipage}[]{0.15\textwidth}
                \epsscale{1.0}
                \plotone{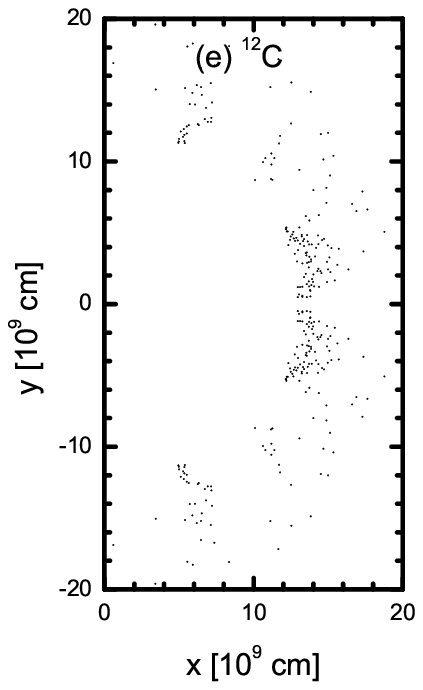}
        \end{minipage}
\end{center}
\caption
{Distribution of characteristic elements/isotopes in the C-DDT model, 
at 10 seconds after the ignition of the deflagration wave. See the caption of Fig.~1. 
\label{fig:fig2}}
\end{figure*}

\begin{figure*}
\begin{center}
        \begin{minipage}[]{0.15\textwidth}
                \epsscale{1.0}
                \plotone{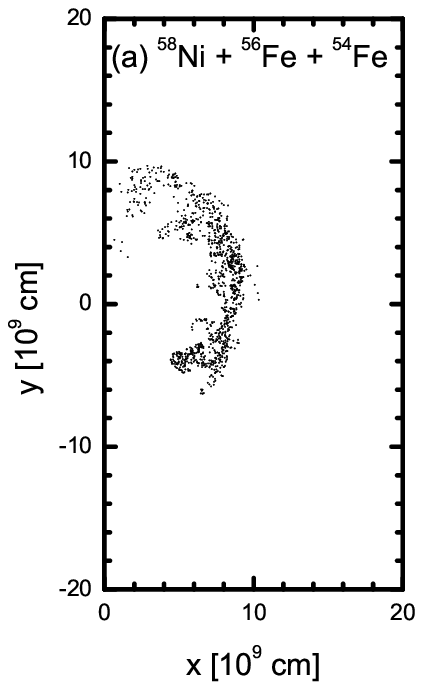}
        \end{minipage}
       \begin{minipage}[]{0.15\textwidth}
                \epsscale{1.0}
                \plotone{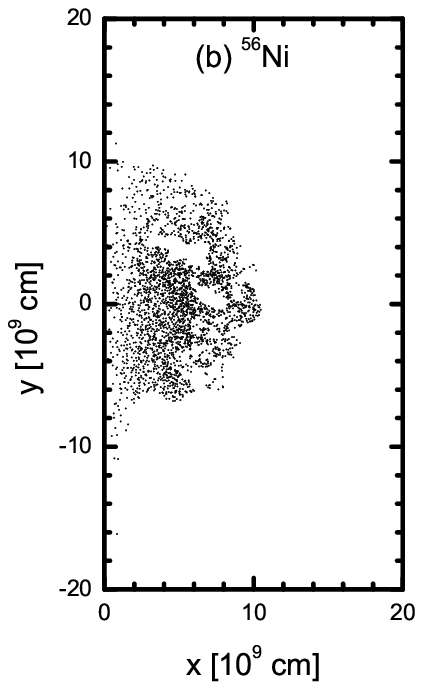}
        \end{minipage}
       \begin{minipage}[]{0.15\textwidth}
                \epsscale{1.0}
                \plotone{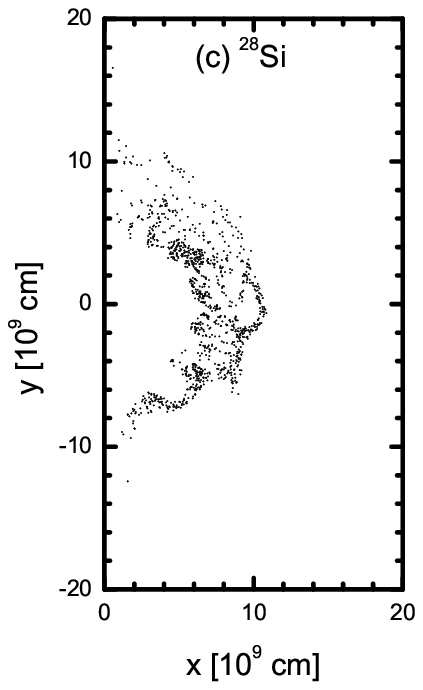}
        \end{minipage}
       \begin{minipage}[]{0.15\textwidth}
                \epsscale{1.0}
                \plotone{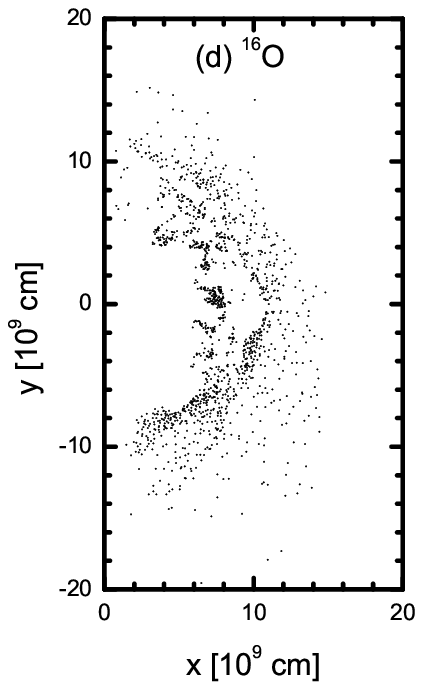}
        \end{minipage}
       \begin{minipage}[]{0.15\textwidth}
                \epsscale{1.0}
                \plotone{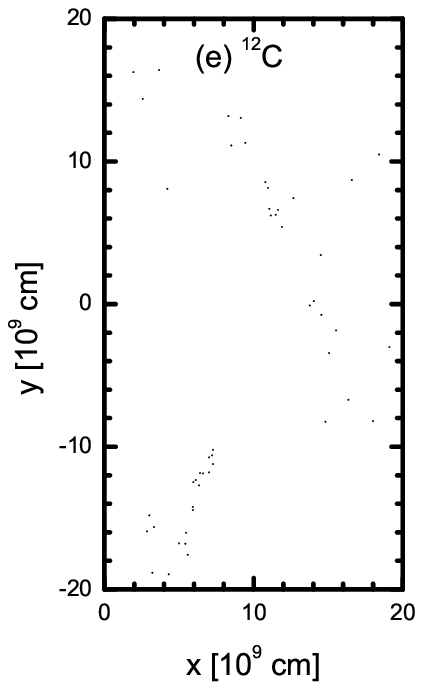}
        \end{minipage}
\end{center}
\caption
{Distribution of characteristic elements/isotopes in the O-DDT model, 
at 10 seconds after the ignition of the deflagration wave. See the caption of Fig.~1. 
\label{fig:fig3}}
\end{figure*}

Throughout this paper, the initial C+O WD composition is assumed as follows: 
X($^{12}$C) $= 0.475$, X($^{16}$O) $=0.5$, and X($^{22}$Ne) $= 0.025$ in mass fractions. 
This is consistent with the compositions used in the W7 model, roughly corresponding to 
the initially solar CNO composition: Metallicity is represented by $^{22}$Ne, assuming that 
the CNO cycle in the H-burning has converted all heavy elements to $^{14}$N, 
and then it is reprocessed 
to $^{22}$Ne by $^{14}$N($\alpha, \gamma$)$^{18}$F(${\rm e}^{+}, \nu_{\rm e}$)
$^{18}$O($\alpha, \gamma$)$^{22}$Ne in the He-burning. 

The reaction network (Thielemann et al.\ 1996) 
includes 384 isotopes up to $^{98}$Mo. 
The electron capture rates, which strongly affect nucleosynthesis 
at the beginning of the deflagration stage, 
are taken from Langanke \& Martinez-Pinedo (2000) and 
Martinez-Pinedo et al. (2000). 
Detailed comparisons between the new rates and 
those of Fuller et al. (1982, 1985) are presented by 
Brachwitz et al. (2000) for the spherical 1D deflagration model W7 
(see also Thielemann et al.\ 2004). 

\begin{figure*}
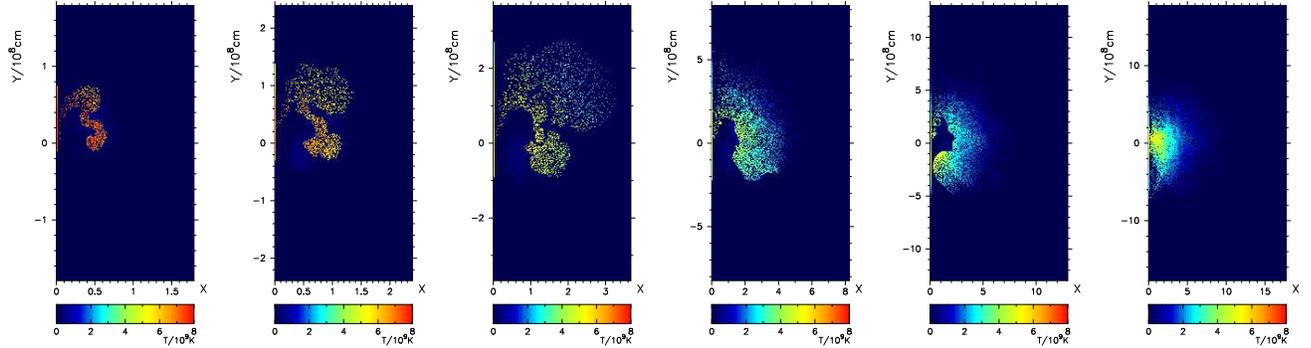

\begin{center}
        \vspace{1cm}
        \hspace{-2cm}
        \begin{minipage}[]{0.15\textwidth}
                \epsscale{2.0}
                \plotone{f4a.eps}
        \end{minipage}
       \begin{minipage}[]{0.15\textwidth}
                \epsscale{2.0}
                \plotone{f4b.eps}
        \end{minipage}
       \begin{minipage}[]{0.15\textwidth}
                \epsscale{2.0}
                \plotone{f4c.eps}
        \end{minipage}
       \begin{minipage}[]{0.15\textwidth}
                \epsscale{2.0}
                \plotone{f4d.eps}
        \end{minipage}
       \begin{minipage}[]{0.15\textwidth}
                \epsscale{2.0}
                \plotone{f4e.eps}
        \end{minipage}
        \begin{minipage}[]{0.15\textwidth}
                \epsscale{2.0}
                \plotone{f4f.eps}
        \end{minipage}
\end{center}
\vspace{1cm}
\caption
{Temporal evolution of temperature distribution (at 0.7, 0.9, 1.1, 1.3, 1.5, 1.7 
seconds after the ignition of the deflagration flame, from left to right), 
in the O-DDT model. The color coordinate corresponds to $T_{9} = 8$ (red) 
and $0$ (blue). The complete silicon burning takes place at the temperature above 
$T_{9} = 5$ (yellow).  
\label{fig:fig4}}
\end{figure*}

\begin{figure*}
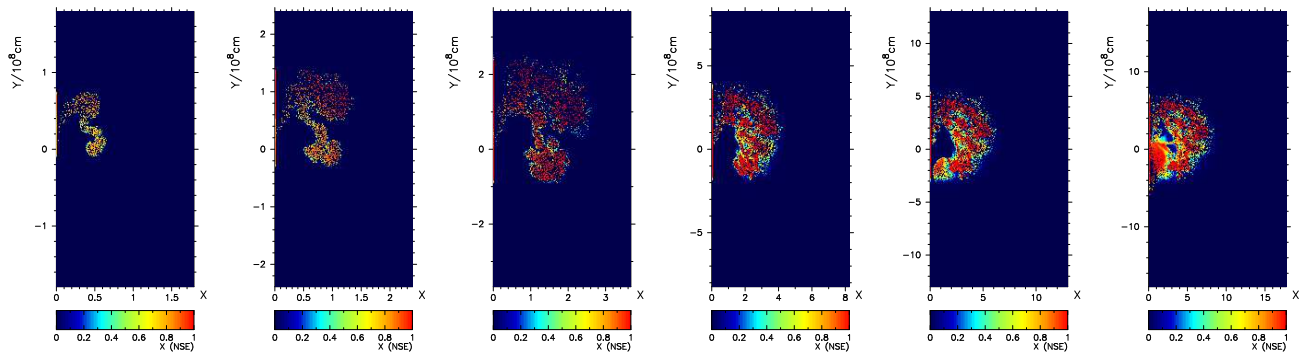

\begin{center}
        \vspace{1cm}
        \hspace{-2cm}
        \begin{minipage}[]{0.15\textwidth}
                \epsscale{2.0}
                \plotone{f5a.eps}
        \end{minipage}
       \begin{minipage}[]{0.15\textwidth}
                \epsscale{2.0}
                \plotone{f5b.eps}
        \end{minipage}
       \begin{minipage}[]{0.15\textwidth}
                \epsscale{2.0}
                \plotone{f5c.eps}
        \end{minipage}
       \begin{minipage}[]{0.15\textwidth}
                \epsscale{2.0}
                \plotone{f5d.eps}
        \end{minipage}
       \begin{minipage}[]{0.15\textwidth}
                \epsscale{2.0}
                \plotone{f5e.eps}
        \end{minipage}
        \begin{minipage}[]{0.15\textwidth}
                \epsscale{2.0}
                \plotone{f5f.eps}
        \end{minipage}
\end{center}
\vspace{1cm}
\caption
{Temporal evolution of distribution of NSE elements (at 0.7, 0.9, 1.1, 1.3, 1.5, 1.7 
seconds after the ignition of the deflagration flame, from left to right), 
in the O-DDT model. Shown here is the mass fraction, from 0 (blue) 
to 1 (red). 
\label{fig:fig5}}
\end{figure*}

\section{Results}

\subsection{Characteristic Burning Regimes}

Before presenting the result of our calculations, 
we summarize the basics of explosive burning taking place 
in the deflagration and detonation stages. 
Typical burning products can be characterized 
by the maximum temperatures ($T_{\rm max}$) and densities 
($\rho_{\rm max}$) attained by 
the material under consideration, after the passage of the 
thermonuclear flame [see, e.g., Arnett (1996), 
Thielemann et al. (1998), and references therein, 
for a review of explosive nucleosynthesis]. 

The explosive burning in thermonuclear supernovae proceeds in
different regimes, mainly characterized by the temperatures reached in
the nuclear ashes (Thielemann et al.\ 1986). In
deflagrations, this directly translates into characteristic fuel
densities ahead of the flame. At the first stage of the deflagration, 
the density is higher than $10^{8}$ g cm$^{-3}$.  
Temperatures rise to $T_{{\rm max}, 9} \equiv 
T_{\rm max}/10^{9}$K $\gsim 6$. At this temperature, 
NSE applies, and thus the final composition is 
determined by freezeout processes (represented 
by $T_{\rm max}$ and $\rho_{\rm max}$) and the efficiency of electron capture reactions. 
Because of the high density, electron capture reactions are important. 
Dominant species in this ``complete silicon burning with electron
capture region''  
are stable $^{56}$Fe, $^{54}$Fe, $^{58}$Ni, and radioactive $^{56}$Ni 
(which decays into $^{56}$Fe). 
In the case of even stronger electron capture reactions, 
the main products are $^{50}$Ti, $^{54}$Cr, and $^{58}$Fe 
(e.g., Thielemann et al.\ 2004). 

Following the expansion of the WD, the deflagration proceeds 
progressively at the lower density. At $T_{{\rm max}, 9} 
\sim 5 - 6$ and $\rho_{\rm max} \sim 5 \times 10^{7} - 10^{8}$ 
g cm$^{-3}$, NSE still applies. 
The electron captures are no longer important, and thus the initial 
$Y_{\rm e}$ is virtually preserved. The dominant species are 
$^{56}$Ni and $^{58}$Ni. 

In contrast to spherically-symmetric delayed detonation models, the
detonation in our two-dimensional setups is triggered at certain spots at
the flame front, but at other locations, deflagration burning can
still proceed for some while.
It is characterized by successively lower temperature 
and density. At $T_{{\rm max}, 9} \sim 3 - 5$, oxygen burning or 
incomplete silicon burning is a result, 
leaving mainly intermediate-mass-elements (IME) such as $^{28}$Si and $^{32}$S. 
At $T_{{\rm max}, 9} \sim 2 - 3$, carbon burning or neon burning is 
a result, characterized by abundant $^{16}$O and $^{24}$Mg 
in consumption of $^{12}$C. At even lower temperatures ($T_{{\rm max}, 9} \lsim 2$), 
virtually no major thermonuclear reactions take place. 

The detonation wave can also be characterized by the same tendency,
although the temperature reached in the ashes is not a unique function
of the fuel density anymore but also depends on the shock strength, the effect 
of which needs further investigation in hydrodynamic simulations. 
In the delayed detonation framework, the detonation is ignited
after significant expansion of 
the WD, and thus the electron captures are almost always unimportant (\S 3.4). If
there is still a high density region left after the  
deflagration, the detonation wave can convert the material chiefly into 
$^{56}$Ni by complete silicon burning. As the density decreases, 
the detonation nucleosynthesis is characterized by oxygen burning, 
carbon burning, and eventually burning ceases. Compared to the
corresponding deflagration case at a similar burning stage, these stages are 
usually encountered at lower fuel
densities, as the detonation compresses the burning material.

\subsection{Structure of the ejecta} 

Table 1 shows the global features of our calculations. 
In the C-DEF model, more than half of the WD material is 
left unburned. The final kinetic energy, $\sim 4 \times 10^{50}$ erg, 
is significantly lower than 
that inferred for normal SNe Ia. In the C-DDT model, 
$\sim 0.5 M_{\odot}$ of the C/O material is incinerated by 
the detonation wave, after the deflagration already has burned 
$\sim 0.5 M_{\odot}$. 
The incinerated mass and nuclear energy release in the deflagration 
stage of the O-DDT model are smaller than the C-DDT model. 
The detonation wave, however, burns a larger amount of material
and produces a larger amount of nuclear energy in the O-DDT model 
than in the C-DDT model. 
The final incinerated mass and the kinetic energy in the O-DDT model 
are comparable to those in the W7 model. 
Hereafter, we discuss 
how these results can be understood in terms of the 
flame propagation and the DDT. 
Note that the nuclear energy release in the hydrodynimic simulations 
(with a simplified treatment for the nuclear reactions) and 
that in the detailed reaction network calculations are 
consistent within 5\%. 

\begin{figure*}
\begin{center}
        \begin{minipage}[]{0.3\textwidth}
                \epsscale{1.0}
                \plotone{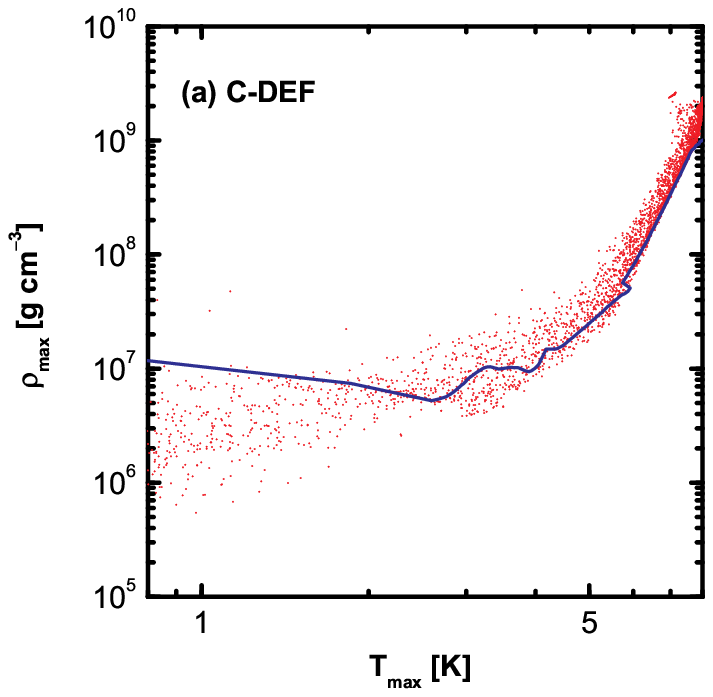}
        \end{minipage}
       \begin{minipage}[]{0.3\textwidth}
                \epsscale{1.0}
                \plotone{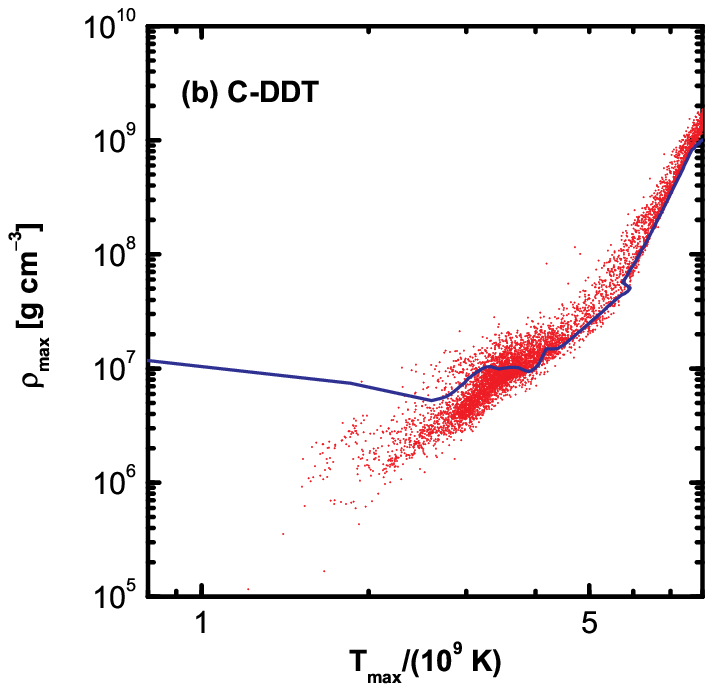}
        \end{minipage}
       \begin{minipage}[]{0.3\textwidth}
                \epsscale{1.0}
                \plotone{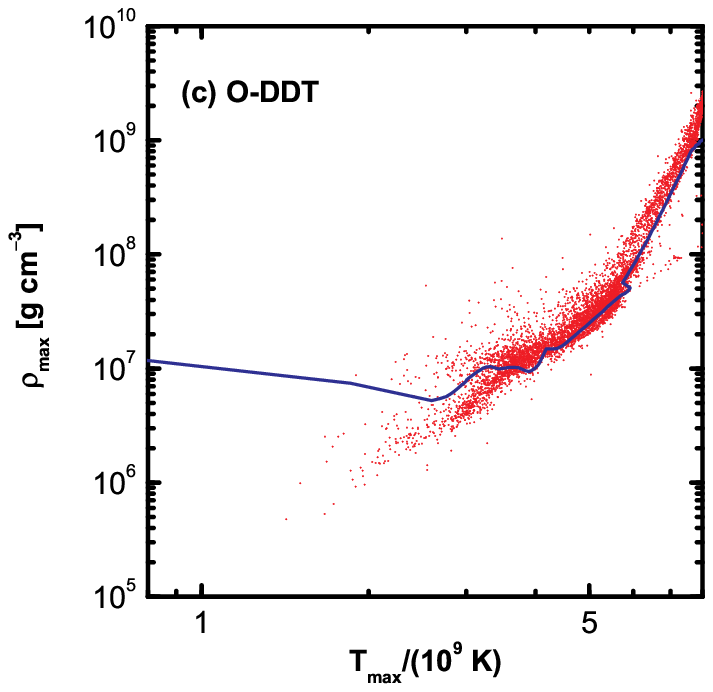}
        \end{minipage}
\end{center}
\caption
{Maximum density $\rho_{\rm max}$ and temperature $T_{\rm max}$ for individual 
tracer particles (dots), after the passage of the thermonuclear flame. 
The solid line is for the W7 model. 
\label{fig:fig6}}
\end{figure*}

Figures 1 -- 3 show the distribution of selected species 
at 10 seconds after the ignition of the deflagration flame.  
At this time, burning has ceased and the SN material is already
almost in a homologous expansion.  
The C-DEF model (Fig.~1) shows large-scale mixing, 
and thus it does not possess a clearly layered structure as predicted in 
exactly spherical 1D models (e.g., W7). A large amount of unburned 
carbon and oxygen are left, being mixed down to the central region. 

In contrast, the detonation wave produces a more layered 
structure, since this supersonically propagating wave 
is unaffected by hydrodynamic instabilities (at least on the large
scales resolved in the models). 
In the C-DDT model, the deflagration burns out much of the center of
the WD before the detonation is triggered.
Therefore, the detonation mainly burns material towards the stellar surface, but  
it also propagates down the fuel funnels between the fingers 
of the deflagration ash (Fig.~1). 
In the C-DDT model, the whole WD, including the central 
region, experiences the strong expansion in the vigorous deflagration stage. 
Therefore, the unburned material left after the deflagration is
processed in the detonation stage mainly by 
carbon and oxygen burning, producing O and IME, but virtually no
Fe-peak elements (Fig.~2). In this particular model,
Fe-peak elements are thus produced mostly in the initial deflagration
stage.

In the O-DDT model the burning proceeds in an aspherical manner. 
Figures 4 and 5 show the temporal evolution
in the O-DDT model. 
The deflagration frame, ignited off-center, 
floats outward, and spreads laterally (R\"opke et al.\ 2007d). 
It creates a large blob of ashes in the upper hemisphere of the
WD\@. Due to the lateral expansion in the outer layers of the star, the
neutron-rich Fe-peak elements produced by  
electron captures end up in a characteristic off-center shell-like 
layer (Fig.~3). 
The detonation triggers on top of this blob at $\sim$ 1 second. 
It cannot cross ash
regions (Maier \& Niemeyer 2006) and has to burn around the ash blob
in order to reach the central parts of the WD\@. Consequently,
it initially propagates only outward (at $\sim$ 1.1 second).

At about 1.5 sec after the deflagration ignition, the detonation wave has
burned around the ash blob and propagates inward. Since the
off-center deflagration did not release much energy, the central parts of the
WD are
still dense and contain mostly unburned material.  
This material is converted to Fe-peak (predominantly $^{56}$Ni) 
by the detonation wave.

\subsection{Characteristic thermal properties}

To understand the characteristic structure of the ejecta described in \S 3.2, 
we examine details of the thermal history of the tracer particles. 
Figure 6 shows the maximum temperature ($T_{\rm max}$) and 
density ($\rho_{\rm max}$) for each 
tracer particle (\S 3.1) after flame passage. For comparison, the same values for the classical 
W7 model are also shown. In Figure 7, average behaviors are plotted,
which were obtained by taking the mass-average as a function of
$T_{\rm max}$. 

Figures 6 and 7 show that the distribution of ($T_{\rm max}$, $\rho_{\rm max}$) 
is similar for all the models 
at $T_{{\rm max}, 9} \gsim 3$ (corresponding to $\rho_{\rm max} 
\sim \rho_{\rm DDT}$). This reflects the fact that the 
local properties of the deflagration flame are basically 
independent from the geometry. 
Interestingly, despite the  different prescription of the 
deflagration flame propagation and different expansion time scale 
(see e.g., Travaglio et al.\ 2004a), we note a similarity with the
outcome of the W7 in this respect. 

\begin{figure}
\begin{center}
        \begin{minipage}[]{0.4\textwidth}
                \epsscale{1.0}
                \plotone{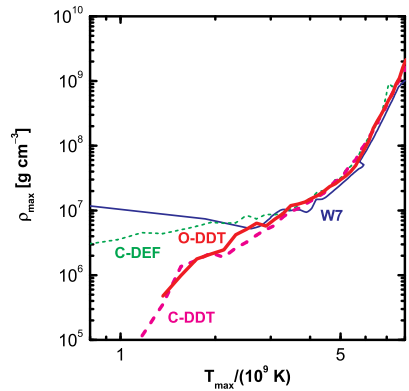}
        \end{minipage}
\end{center}
\caption
{Average peak density of tracer particles as a function of peak temperature 
(with the temperature bin taken to be $\Delta T_{\rm max} = 2 \times 10^{8}$ K). 
Models are W7 (blue thin-solid), 
C-DEF (green thin-dashed), C-DDT (magenta thick-dashed), 
and O-DDT (red thick-solid). 
\label{fig:fig7}}
\end{figure}

\begin{figure}
\begin{center}
        \begin{minipage}[]{0.4\textwidth}
                \epsscale{1.0}
                \plotone{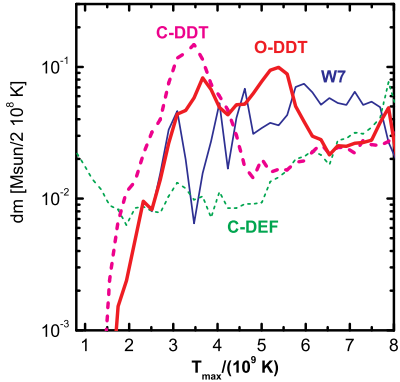}
        \end{minipage}
\end{center}
\caption
{The masses (\Msun) within each peak temperature range 
($\Delta T = 2 \times 10^8$ K). Models are W7 (blue thin-solid), 
C-DEF (green thin-dashed), C-DDT (magenta thick-dashed), 
and O-DDT (red thick-solid). 
\label{fig:fig8}}
\end{figure}

At $T_{{\rm max}, 9} \lsim 3$, the models differ. 
Note that the behavior of W7 in this temperature regime shown in Figures 6 and 7 
is not real; the deflagration is turned off at $T_{{\rm max}, 9} \sim 2$, 
where the flame is already near the surface in the W7 model. 
Our deflagration model extends to 
lower $T_{\rm max}$ and $\rho_{\rm max}$. In the two delayed detonation models, 
this temperature range corresponds to the detonation phase, and thus the behavior 
is expected to deviate from the deflagration models, and indeed it does. 
The two detonation models show a similar $T_{\rm max} - \rho_{\rm max}$ distribution; 
the compression and the maximum temperature are roughly linear functions 
of the unburned density. 

Figure 6 shows a significant dispersion in the $T_{\rm max} - \rho_{\rm max}$ distribution 
around the mean value, reaching nearly one order of magnitude 
in $\rho_{\rm max}$ for given $T_{\rm max}$, at $T_{{\rm max}, 9} \lsim 5$. 
This is a unique behavior in the multi-D models, potentially 
leading to a variety of burning products otherwise not expected in strictly 
spherically symmetric models like W7. 

The different hydrodynamic properties lead to different 
amounts of material as a function of $T_{\rm max}$ (or $\rho_{\rm max}$), 
and this is the main reason why different nucleosynthesis features appear. 
Figure 8 shows the mass of the material as a function of $T_{\rm max}$. 
It is seen that the deflagration 
is efficient in W7, and thus the W7 model has a much larger amount of the 
mass with $T_{{\rm max}, 9} \gsim 6$ than the 
2D models. At $T_{{\rm max}, 9} \lsim 3 - 4$, there appears a 
peak in the C-DDT model; this is due to the detonation wave 
burning the fuel material left behind the deflagration stage. 
The O-DDT model shows a characteristic distribution; two peaks 
at $T_{{\rm max}, 9} \sim 3 - 4$ and at $T_{{\rm max}, 9} \sim 5 - 6$. 
The former is the detonation wave propagating outward, 
as also seen in the C-DDT model. The latter one 
at higher $T_{\rm max}$ is due to the detonation wave propagating inward, 
near the central region where the fuel still has high densities. 
The gap between the two peaks corresponds to the (off-center) shell structure 
of the deflagration products (i.e., neutron-rich Fe-peak elements; Fig.~3). 

\begin{figure}
\begin{center}
        \begin{minipage}[]{0.4\textwidth}
                \epsscale{1.0}
                \plotone{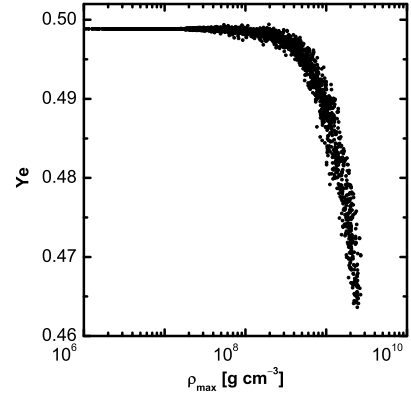}
        \end{minipage}
\end{center}
\caption
{The final values of $Y_{\rm e}$ as a function of the peak 
density, for the O-DDT model.  
\label{fig:fig9}}
\end{figure}

\subsection{Deflagration and Electron Captures}

In the densest region in the deflagration stage, 
electron capture reactions proceed to synthesize neutron-rich 
Fe-peak elements under NSE conditions. Figure 9 shows 
the electron fraction ($Y_{\rm e}$) as a function of 
$\rho_{\rm max}$, for the O-DDT model. 
The same figure for the other 2D models (not shown) is similar to 
the O-DDT model. Figure 9 shows that 
the electron capture becomes important at $\rho_{\rm max} 
\gsim 5 \times 10^{8}$ g cm$^{-3}$, and $Y_{\rm e}$ 
as low as $0.463$ is realized in the highest density region. 
Thus, the main product of the highest density region is 
stable $^{56}$Fe ($Y_{\rm e} \sim 0.464$), 
followed by $^{54}$Fe ($Y_{\rm e} \sim 0.481$) 
and $^{58}$Ni ($Y_{\rm e} \sim 0.483$) 
in the lower density region. 
Since $Y_{\rm e} > 0.46$, $^{50}$Ti ($Y_{\rm e} \sim 0.440$), 
$^{54}$Cr ($Y_{\rm e} \sim 0.444$) 
and $^{58}$Fe ($Y_{\rm e} \sim 0.448$) are not produced abundantly 
in the present models. 
The behavior in $Y_{\rm e}$ and the resulting electron capture reactions are 
largely consistent with the 2D deflagration model 
presented in Travaglio et al. (2004a). The electron captures, however, 
are less efficient in the 2D models than in the 3D deflagration model 
(Travaglio et al.\ 2004a) and in the classical 1D models (Thielemann et al.\ 2004; 
see also \S 3.6). 
This is likely due to the weaker deflagration in the present models than in 
the others (1D and 3D), although an uncertainty is involved in the treatment of the 
temperature evolution in the NSE regime (\S 2.2). 

Figure 10 shows the comparison between $Y_{\rm e}$ before and after the DDT. 
In the C-DDT model, $Y_{\rm e}$ is not affected by the detonation wave as is 
consistent with previous 1D models. On the other hand, in the O-DDT model 
the detonation slightly affects $Y_{\rm e}$ especially of the material which was 
hardly processed 
by the deflagration ($\sim Y_{\rm e} \gsim 0.496$). The change in $Y_{\rm e}$ 
introduced by the detonation is at most $\Delta Y_{\rm e} \sim 0.02$, much smaller 
than the change in the deflagration stage at $\gsim 10^{8} - 10^{9}$g cm$^{-3}$. 
This is a result of electron capture reactions at the detonation wave propagating 
near the WD center, since the central region is still at high density (e.g., Meakin et al. 2009). 
The change is not as large as that in the GCD models of Jordan et al. (2008) and 
Meakin et al. (2009), which resulted in $\Delta Y_{\rm e} \sim 0.05$. This is because 
the distribution of the initial bubbles 
in our simulations is not as extreme as theirs (see \S 2.1) and 
the WD pre-expansion before the onset of the detonation was stronger
in our model.

\begin{figure*}
\begin{center}
        \begin{minipage}[]{0.4\textwidth}
                \epsscale{1.0}
                \plotone{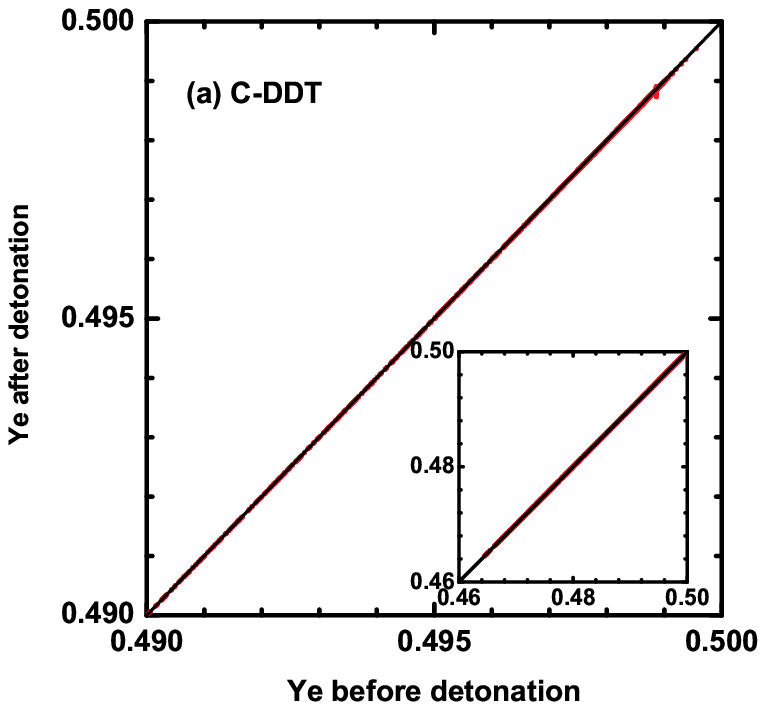}
        \end{minipage}
         \begin{minipage}[]{0.4\textwidth}
                \epsscale{1.0}
                \plotone{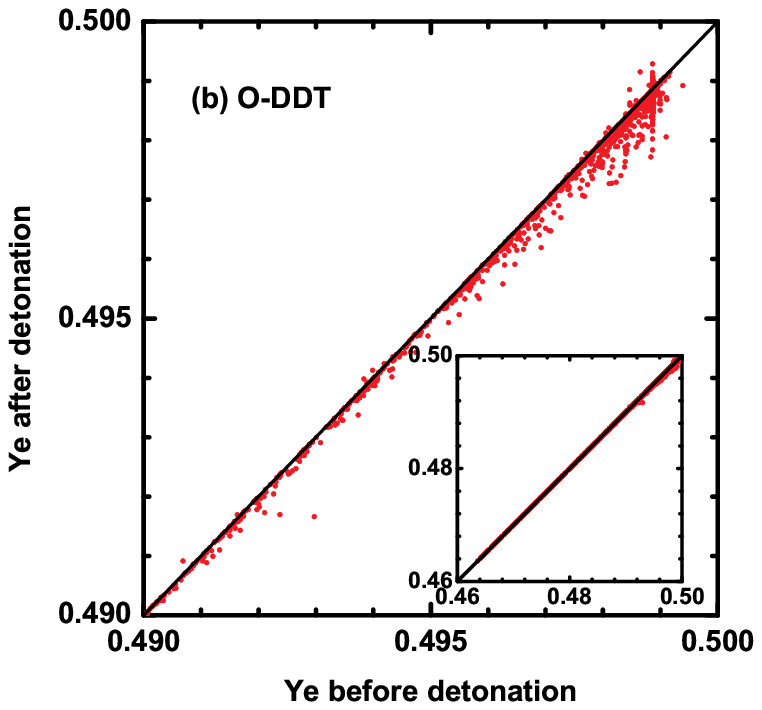}
        \end{minipage}
\end{center}
\caption
{The final values of $Y_{\rm e}$ (i.e., after the detonation) as compared 
to those just before the first DDT transition (i.e., before the detonation), 
for (a) C-DDT and (b) O-DDT models (shown by dots). 
\label{fig:fig10}}
\end{figure*}

\subsection{Distribution of Nucleosynthesis Products}

Figure 11 shows the radial abundance distribution, for which the 
material is angle-averaged within each velocity bin. 
Note that our models are followed up to $\sim 10$ seconds, 
thus the velocity is proportional to the radial distance, 
and the distribution in velocity space corresponds in good
approximation to that in spatial space. The O-DDT model has a very 
aspherical ejecta structure. Therefore, the angle-dependent radial abundance distribution 
in this model is given in Figure 12.

In the C-DEF model,  
different burning regions are macroscopically well-mixed. 
As a consequence, it forms a broad region 
in which the electron capture 
products ($^{56}$Fe, $^{54}$Fe, and $^{58}$Ni), 
complete-silicon burning products ($^{56}$Ni), 
intermediate mass elements ($^{32}$S, $^{28}$Si, $^{24}$Mg), 
and unburned elements ($^{16}$O and $^{12}$C) coexist 
(when averaging in the zenith angle). This region is surrounded by
unburned material at radii not reached by the flame. An enhancement of 
C and O near the center due to the large-scale deflagration mixing is
visible. 
These are typical of 2D deflagration models; this smoothing effect 
tends to be suppressed in 3D deflagration models (R\"opke et al.\ 2007b). 

In the C-DDT model, 
the inner region below $\sim 8,000$ km s$^{-1}$ has a similar 
structure to that in the deflagration model. The difference 
is seen in the innermost region, where unburned C and O are burned into 
intermediate mass elements by the detonation wave. 
In this model, C is almost completely burned, but there still remains 
O with a mass fraction of $\sim 0.1$, since the detonation 
in this model leads to carbon or oxygen burning. 

Surrounding the deflagration region is the oxygen burning layer, 
where $^{28}$Si, $^{32}$S, and $^{24}$Mg are the main products. 
This is surrounded by a mixture of carbon burning products and unburned 
material. Note that the carbon burning region and the unburned region are 
microscopically separated. As shown in Figure 2, the unburned region 
(with almost initial C+O composition) is not fully homogeneously 
distributed, and there are ``fingers'' of carbon burning regions 
(where oxygen and IMEs are produced in consumption of carbon). 

\begin{figure*}
\begin{center}
        \begin{minipage}[]{0.3\textwidth}
                \epsscale{1.0}
                \plotone{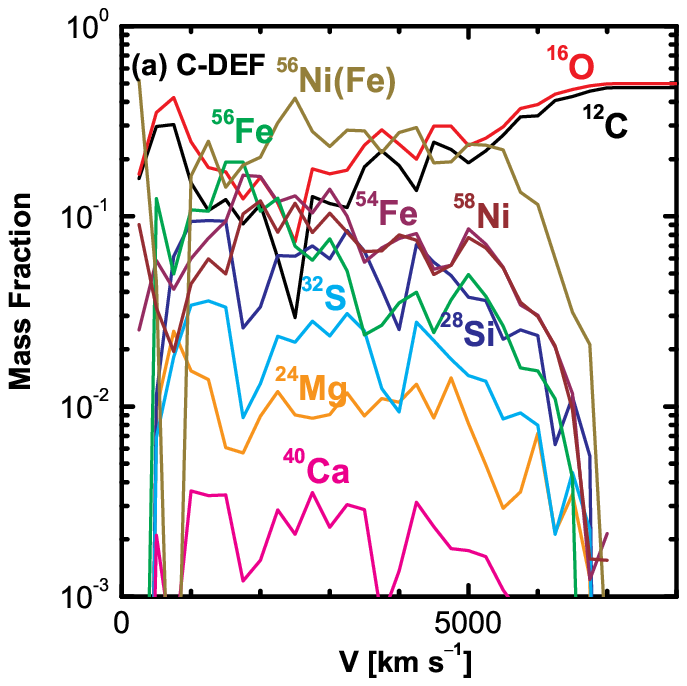}
        \end{minipage}
       \begin{minipage}[]{0.3\textwidth}
                \epsscale{1.0}
                \plotone{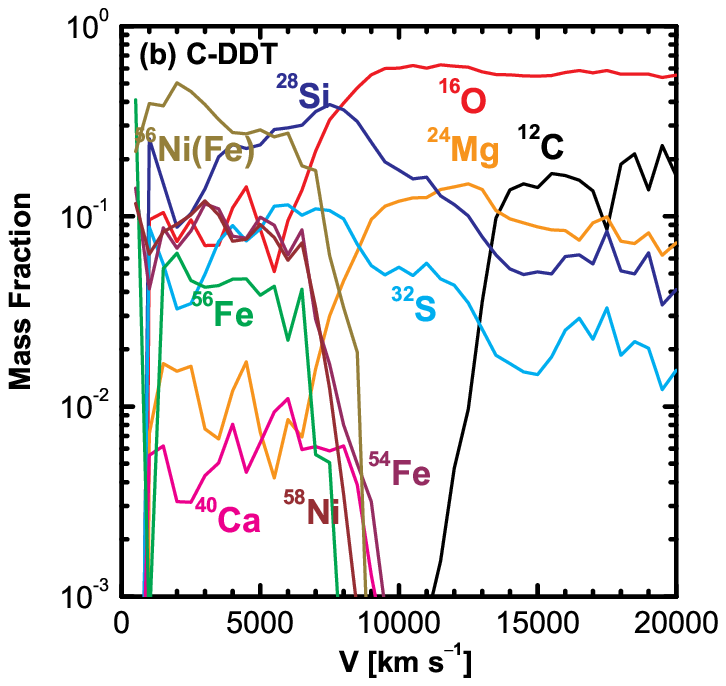}
        \end{minipage}
       \begin{minipage}[]{0.3\textwidth}
                \epsscale{1.0}
                \plotone{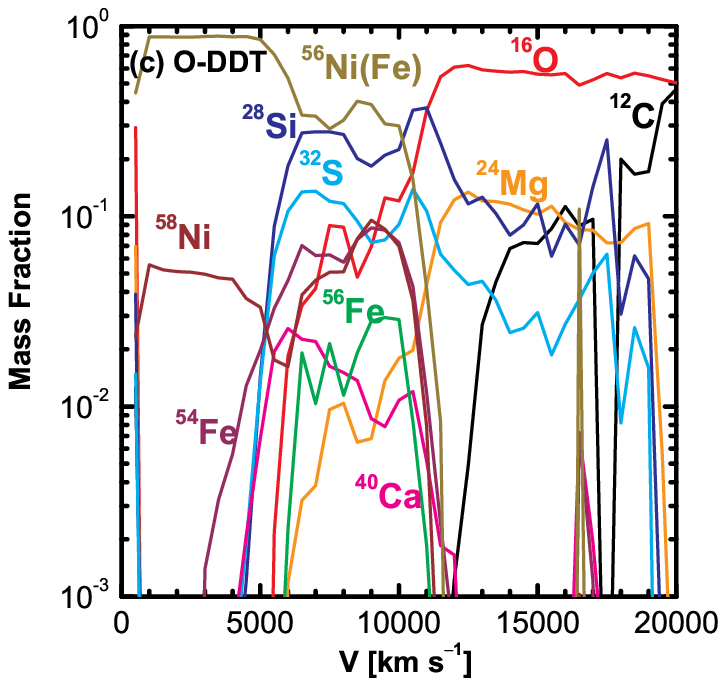}
        \end{minipage}\\
       \begin{minipage}[]{0.3\textwidth}
                \epsscale{1.0}
                \plotone{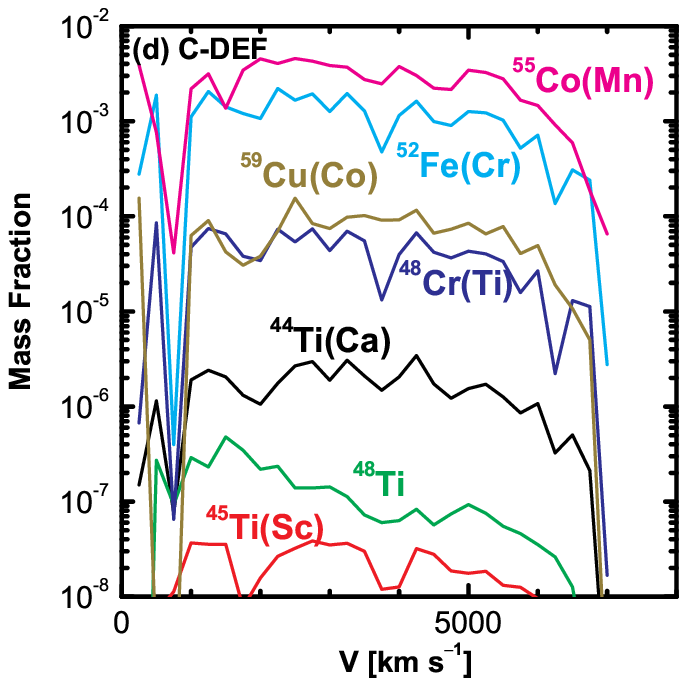}
        \end{minipage}
       \begin{minipage}[]{0.3\textwidth}
                \epsscale{1.0}
                \plotone{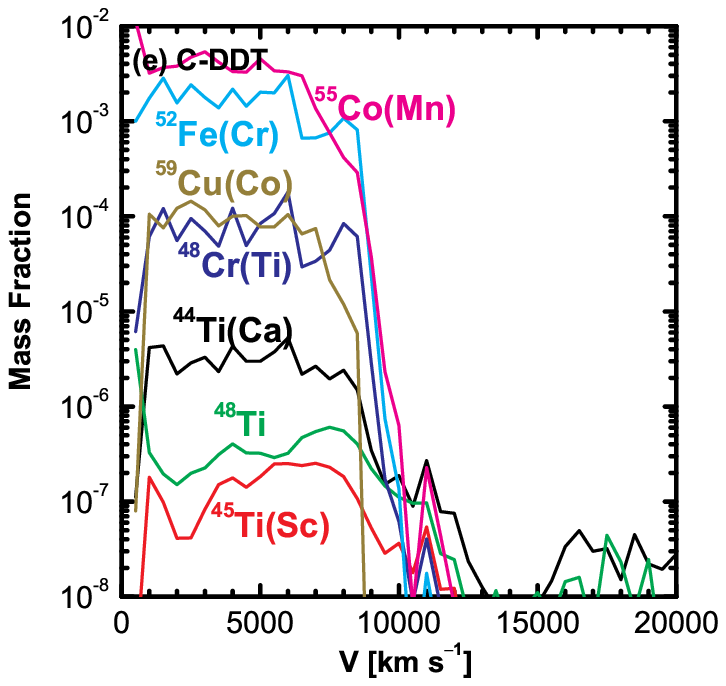}
        \end{minipage}
       \begin{minipage}[]{0.3\textwidth}
                \epsscale{1.0}
                \plotone{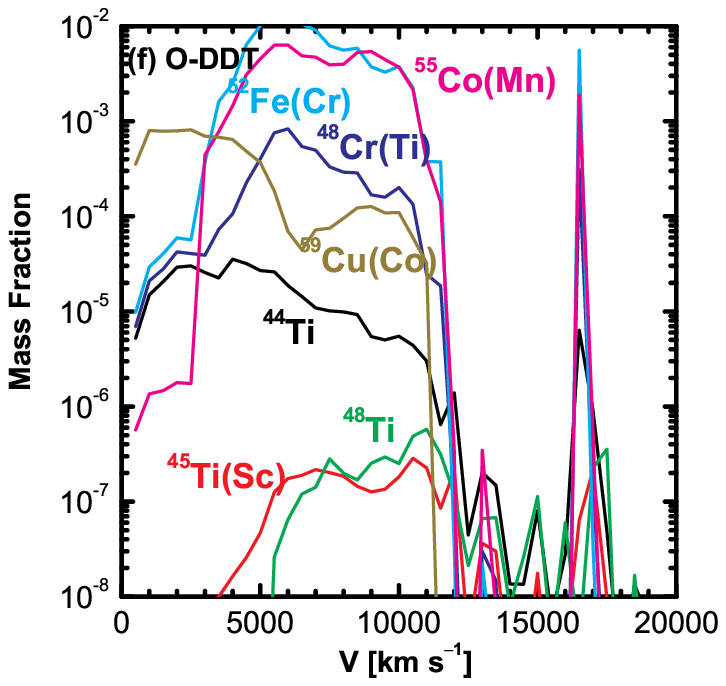}
        \end{minipage}
\end{center}
\caption
{Mass fractions of selected isotopes as a function of 
the expansion velocity, averaged in the velocity bin of 
$\Delta V$. (a, d) C-DEF model, $\Delta V = 250$ km s$^{-1}$, 
(b, e) C-DDT, $\Delta V = 500$ km s$^{-1}$, 
and (c, f) O-DDT, $\Delta V = 500$ km s$^{-1}$. 
\label{fig:fig11}}
\end{figure*}

\begin{figure*}
\begin{center}
        \begin{minipage}[]{0.3\textwidth}
                \epsscale{1.0}
                \plotone{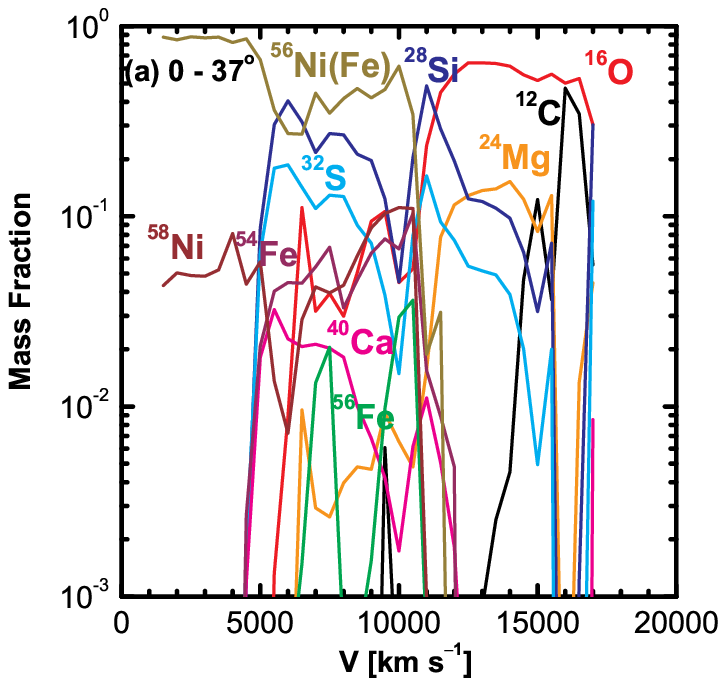}
        \end{minipage}
       \begin{minipage}[]{0.3\textwidth}
                \epsscale{1.0}
                \plotone{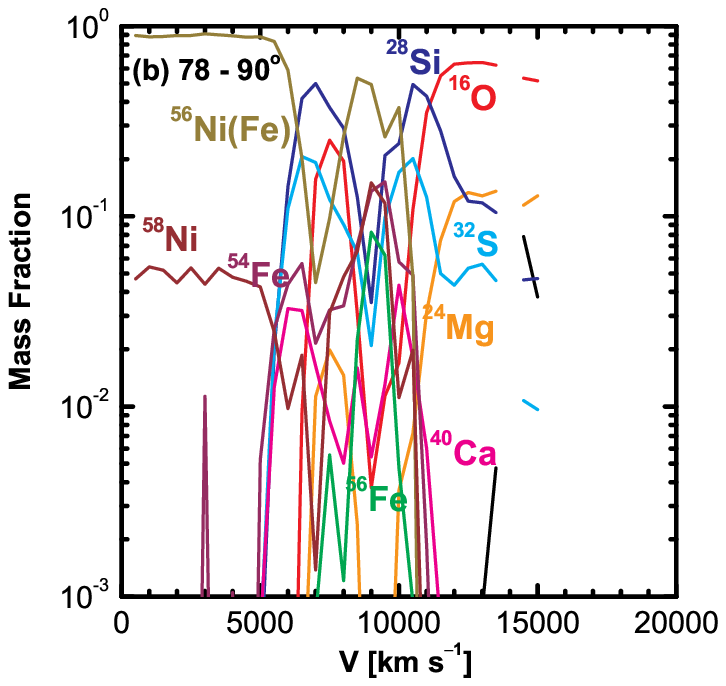}
        \end{minipage}
       \begin{minipage}[]{0.3\textwidth}
                \epsscale{1.0}
                \plotone{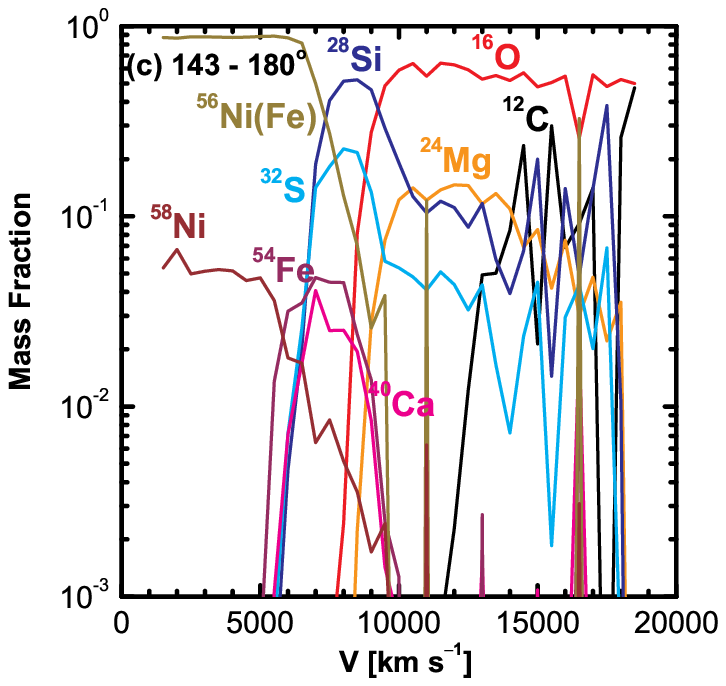}
        \end{minipage}\\
       \begin{minipage}[]{0.3\textwidth}
                \epsscale{1.0}
                \plotone{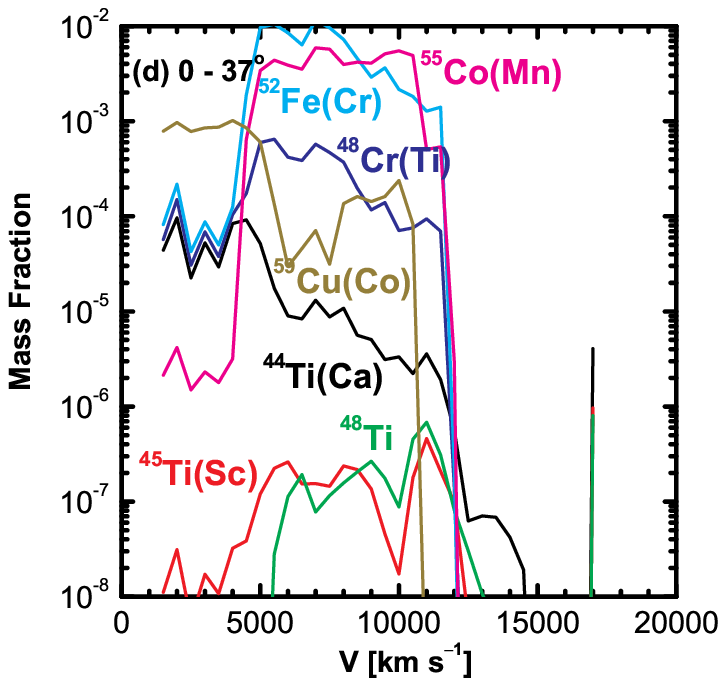}
        \end{minipage}
       \begin{minipage}[]{0.3\textwidth}
                \epsscale{1.0}
                \plotone{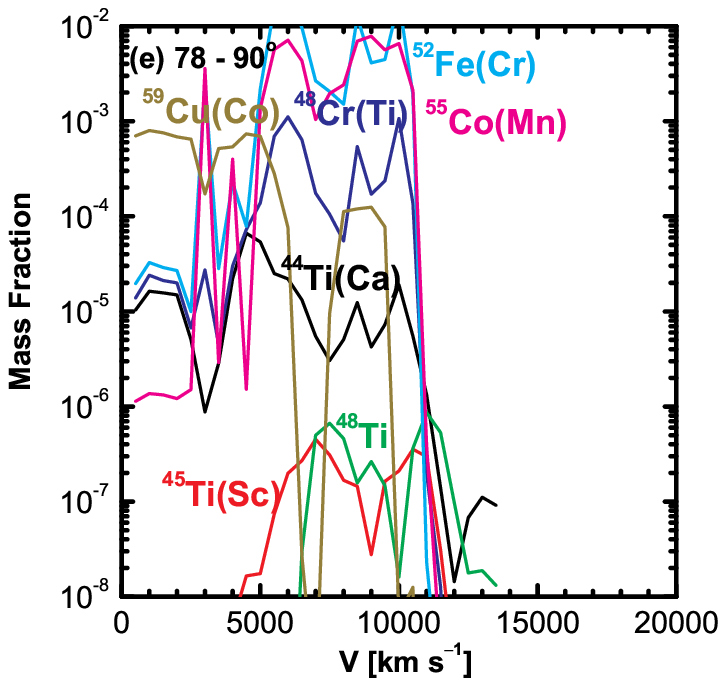}
        \end{minipage}
       \begin{minipage}[]{0.3\textwidth}
                \epsscale{1.0}
                \plotone{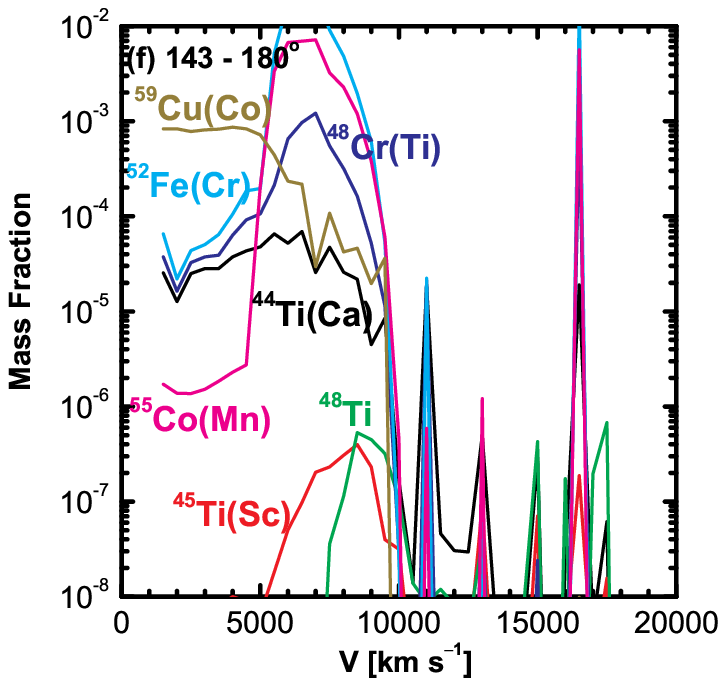}
        \end{minipage}
\end{center}
\caption
{Angle-dependent mass fractions of selected isotopes 
in O-DDT, as a function of 
the expansion velocity, averaged in the velocity bin of 
$\Delta V = 500$ km s$^{-1}$. (a, d) averaged within 0 -- 37 deg,  
(b, e) 78 -- 90 deg, 
and (c, f) 143 -- 180 deg. 
\label{fig:fig12}}
\end{figure*}

The radial composition distribution in the O-DDT model 
follows that of the C-DDT model. A striking difference, however, is
that the region of deflagration ashes is not in the center but confined within 
$V \sim 5,000 - 10,000$ km s$^{-1}$. The innermost region is 
converted to $^{56}$Ni, by the complete silicon burning 
taking place at the inward detonation wave.

\begin{figure*}
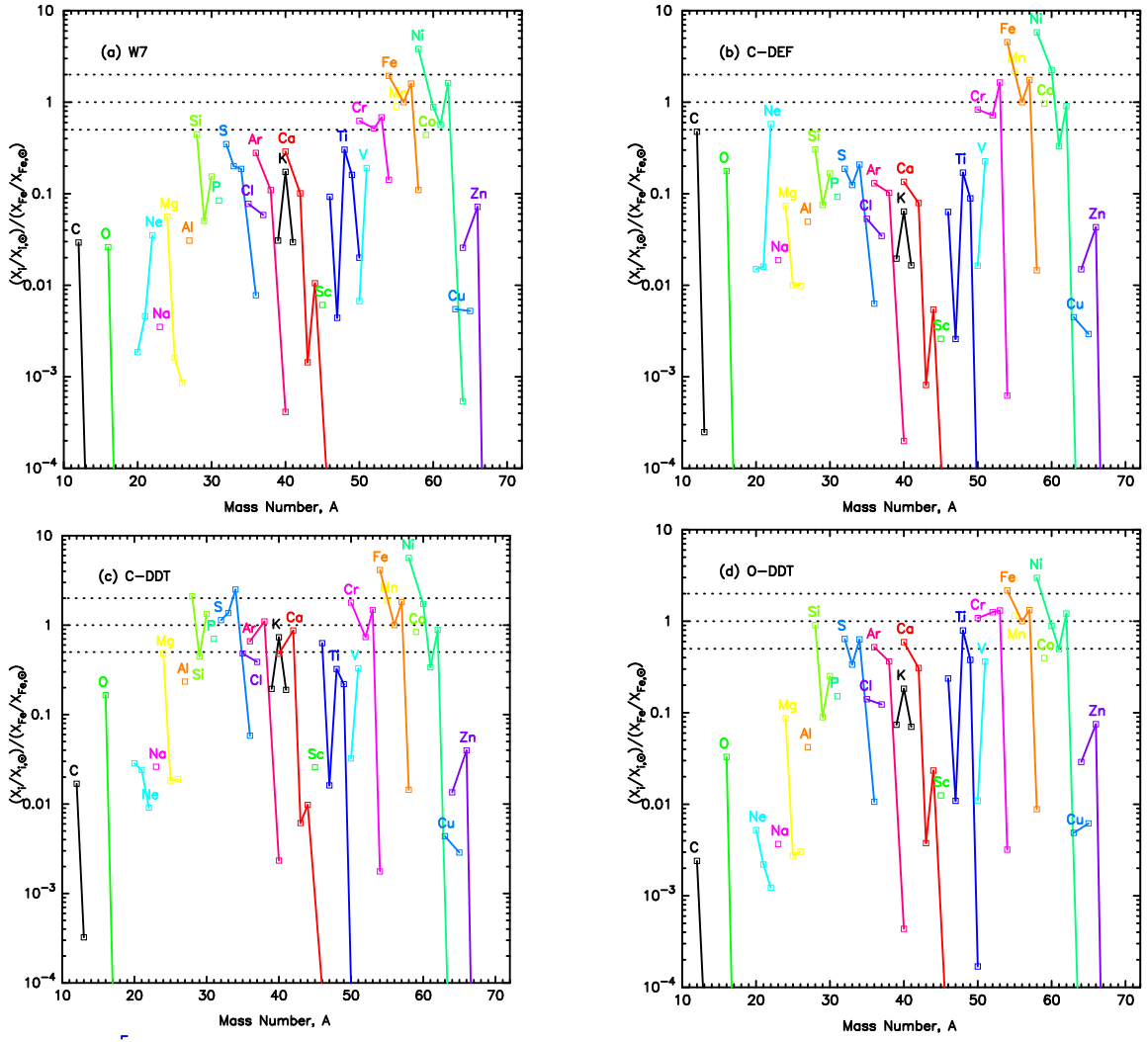

\begin{center}
        \begin{minipage}[]{0.45\textwidth}
                \epsscale{1.0}
                \plotone{f13a.eps}
        \end{minipage}
       \begin{minipage}[]{0.45\textwidth}
                \epsscale{1.0}
                \plotone{f13b.eps}
        \end{minipage}
       \begin{minipage}[]{0.45\textwidth}
                \epsscale{1.0}
                \plotone{f13c.eps}
        \end{minipage}
       \begin{minipage}[]{0.45\textwidth}
                \epsscale{1.0}
                \plotone{f13d.eps}
        \end{minipage}
\end{center}
\caption
{Isotopic yields integrated within the whole ejecta 
(after radioactive decays). 
The mass fraction is shown relative to the solar value, 
and normalized by $^{56}$Fe. 
\label{fig:fig13}}
\end{figure*}

\begin{figure*}
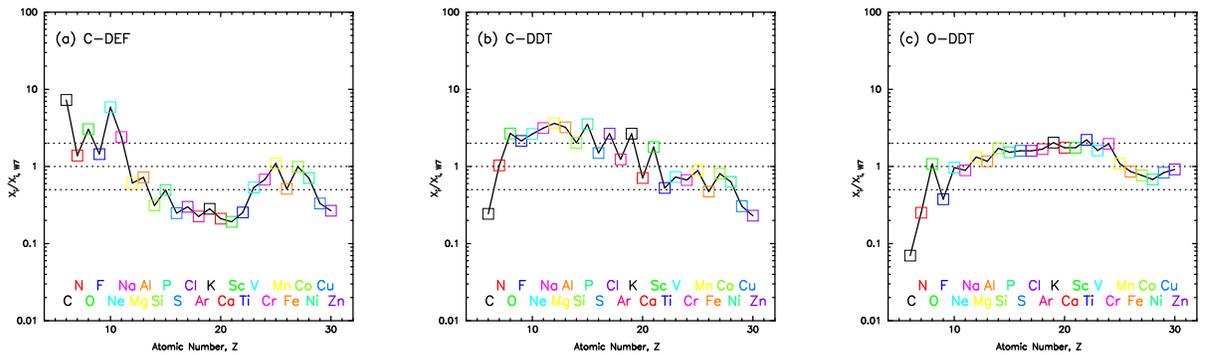

\begin{center}
        \begin{minipage}[]{0.3\textwidth}
                \epsscale{1.0}
                \plotone{f14a.eps}
        \end{minipage}
       \begin{minipage}[]{0.3\textwidth}
                \epsscale{1.0}
                \plotone{f14b.eps}
        \end{minipage}
       \begin{minipage}[]{0.3\textwidth}
                \epsscale{1.0}
                \plotone{f14c.eps}
        \end{minipage}
\end{center}
\caption
{Ratios of integrated element yields to the W7 model yields. 
\label{fig:fig14}}
\end{figure*}

\begin{deluxetable*}{ccccccc}
 \tabletypesize{\scriptsize}
 \tablecaption{Masses of elements (\Msun)
 \label{tab:tab1}}
 \tablewidth{0pt}
 \tablehead{
   \colhead{Element}
 & \colhead{W7}
 & \colhead{C-DEF}
 & \colhead{C-DDT}
 & \colhead{O-DDT}
 & \colhead{C-DDT/def.\tablenotemark{a}}
 & \colhead{O-DDT/def.\tablenotemark{a}}
}
\startdata
C  &  4.99E-02 & 3.64E-01 & 1.21E-02 & 3.49E-03 & 4.17E-01 & 4.86E-01\\
N  &  1.11E-06 & 1.54E-06 & 1.15E-06 & 2.80E-07 & 2.55E-06 & 1.14E-05\\
O  &  1.40E-01 & 4.27E-01 & 3.76E-01 & 1.51E-01 & 4.61E-01 & 5.21E-01\\
F  &  7.19E-10 & 1.04E-09 & 1.54E-09 & 2.70E-10 & 5.40E-10 & 3.56E-10\\
Ne &  4.28E-03 & 2.51E-02 & 1.13E-02 & 4.13E-03 & 2.49E-02 & 2.82E-02\\
Na &  6.59E-05 & 1.59E-04 & 2.07E-04 & 5.87E-05 & 7.09E-05 & 6.15E-05\\
Mg &  1.63E-02 & 9.92E-03 & 5.90E-02 & 2.16E-02 & 4.81E-03 & 2.27E-03\\
Al &  1.00E-03 & 7.23E-04 & 3.22E-03 & 1.17E-03 & 3.45E-04 & 1.99E-04\\
Si &  1.67E-01 & 5.19E-02 & 3.38E-01 & 2.87E-01 & 3.39E-02 & 1.38E-02\\
P  &  3.86E-04 & 1.91E-04 & 1.36E-03 & 5.91E-04 & 1.34E-04 & 8.34E-05\\
S  &  7.97E-02 & 1.98E-02 & 1.19E-01 & 1.27E-01 & 1.33E-02 & 6.05E-03\\
Cl &  1.38E-04 & 4.15E-05 & 3.69E-04 & 2.20E-04 & 5.81E-05 & 7.83E-05\\
Ar &  1.31E-02 & 2.94E-03 & 1.62E-02 & 2.20E-02 & 2.41E-03 & 1.61E-03\\
K  &  6.45E-05 & 1.82E-05 & 1.71E-04 & 1.31E-04 & 3.25E-05 & 8.11E-05\\
Ca &  9.76E-03 & 2.05E-03 & 6.91E-03 & 1.70E-02 & 1.72E-03 & 1.60E-03\\
Sc &  1.33E-07 & 2.55E-08 & 2.38E-07 & 2.29E-07 & 1.59E-07 & 1.59E-05\\
Ti &  3.94E-04 & 9.97E-05 & 2.08E-04 & 8.72E-04 & 1.53E-04 & 1.98E-03\\
V  &  4.04E-05 & 2.15E-05 & 2.95E-05 & 6.52E-05 & 1.29E-04 & 3.01E-03\\
Cr &  5.28E-03 & 3.57E-03 & 3.52E-03 & 1.04E-02 & 4.41E-03 & 2.07E-02\\
Mn &  6.72E-03 & 7.35E-03 & 5.95E-03 & 7.35E-03 & 1.83E-02 & 4.42E-02\\
Fe &  7.61E-01 & 3.89E-01 & 3.60E-01 & 6.51E-01 & 3.36E-01 & 2.06E-01\\
Co &  8.29E-04 & 8.20E-04 & 6.70E-04 & 6.36E-04 & 1.31E-03 & 3.22E-03\\
Ni &  1.19E-01 & 8.42E-02 & 7.50E-02 & 8.05E-02 & 6.88E-02 & 4.30E-02\\
Cu &  2.55E-06 & 8.45E-07 & 7.82E-07 & 2.13E-06 & 1.38E-06 & 1.26E-05\\
Zn &  3.80E-05 & 1.01E-05 & 8.82E-06 & 3.50E-05 & 1.08E-06 & 5.48E-06\\
\enddata
\tablenotetext{a}{At the moment when the first DDT takes place.}
\end{deluxetable*}

\subsection{Total Yields}

Figure 13 shows the nucleosynthesis yield for each model, 
integrated over the whole ejecta. 
The ratios of the masses of elements 
to those of the W7 model are plotted in Figure 14. 
Tables 2 and 3 list the masses 
of the elements and isotopes (after radioactive decays) while major
long-lived radioactive isotopes are given in
Table 4. For the C-DDT and O-DDT models, 
the values at the time of the first DDT ($\sim 1.15$ sec for C-DDT and 
$\sim 1.05$ sec for O-DDT) are also shown 
in Tables 2 and 3. 

For comparison, we present the yields of the W7 model, 
calculated by ourselves using the thermal history of the original model. 
Details are different from Iwamoto et al. (1999), 
because of the updated electron capture rates (see Brachwitz et al.\ 2000). 

The result for the C-DEF model is consistent with 
the similar 2D model in the previous study (Travaglio et al.\ 2004a). 
First of all, the mass of $^{56}$Ni is only $\sim 0.25 M_{\odot}$, 
smaller than in the W7 model ($\sim 0.64 M_{\odot}$). 
A large amount of C and O are left unburned. 
The final C/O ratio 
is larger than in the W7 model, because of the weak C-burning 
in our model resulting in almost the original, unburned C/O ratio near the surface 
(Fig.~11). IMEs are underproduced as compared to the
W7 model. Note, however, that generally full three-dimensional 
models alleviate this problem (e.g., Travaglio et al.\ 2004a; R\"opke et al. 2007b). 

In the C-DDT model, the density of the WD 
is already low when the DDT takes place, and the temperature of material after the 
passage of the detonation wave does not reach $\sim 5 \times 10^{9}$ K 
(Fig.~8). The detonation therefore produces at most IMEs; 
Fe-peak elements are produced in the initial deflagration stage, 
and the mass of $^{56}$Ni is almost the same as in the deflagration model. 
Consequently, the abundance pattern of the Fe-peak elements (including $^{56}$Ni) 
is very similar to that of the C-DEF model (Figs.~13 and 14). 
IMEs are produced much more abundantly than in the W7 model. 
Because of the carbon burning, the C/O ratio is much smaller than for W7. 

Somewhat surprisingly, the O-DDT model predicts the abundance pattern 
similar to W7 (Fig.~13), with most of elements consistent with 
W7 within a factor of two (Fig.~14). 
The mass of $^{56}$Ni ($\sim 0.54 M_{\odot}$) 
is comparable with the W7 model ($\sim 0.64 M_{\odot}$) 
(but see \S 4). 
IMEs tend to be overproduced by a factor of $\sim 2$, because of the 
carbon and oxygen burning in the detonation phase. 
Carbon is almost entirely consumed by the carbon burning. 

A comparison between the W7 model and our 2D models shows that 
very strong electron capture reactions are not efficient in the 
2D models (\S 3.4). $^{50}$Ti, $^{54}$Cr, and $^{58}$Fe, produced at 
the highest density resulting in $Y_{\rm e} < 0.46$, are not abundantly 
produced in the 2D models. 

\subsection{Radioactive Isotopes}

The amount of some radioactive isotopes in our detonation 
models is quite different from W7 and even with canonical, spherical 
delayed detonation models. Comparison between our results (Tab.~4) and Table~4 of 
Iwamoto et al. (1999) shows that radioactive isotopes lighter than Fe-peak 
tend to be more abundant, with the tendency of  
larger amounts of these isotopes for smaller density at the detonation. 
On the other hand, the amount of some Fe-peak radioactive isotopes can change 
dramatically from the classical 1D delayed detonation model, due to updated 
electron capture rates in our calculations (\S 2.2).

\begin{deluxetable*}{ccccccc}
 \tabletypesize{\scriptsize}
 \tablecaption{Masses of isotopes (\Msun)
 \label{tab:tab2}}
 \tablewidth{0pt}
 \tablehead{
   \colhead{Isotopes}
 & \colhead{W7}
 & \colhead{C-DEF}
 & \colhead{C-DDT}
 & \colhead{O-DDT}
 & \colhead{C-DDT / def.}
 & \colhead{O-DDT / def.}
}
\startdata
 $^{12}$C &  4.99E-02  & 3.64E-01  & 1.21E-02  & 3.49E-03  & 4.17E-01  & 4.86E-01\\
 $^{13}$C  & 9.57E-07  & 2.29E-06  & 2.80E-06  & 8.75E-07  & 1.16E-06  & 1.35E-06\\
 $^{14}$N  & 1.11E-06  & 1.54E-06  & 1.15E-06  & 2.79E-07  & 2.55E-06  & 1.14E-05\\
 $^{15}$N  & 1.71E-09  & 1.65E-09  & 3.32E-09  & 7.06E-10  & 2.79E-09  & 9.71E-09\\
 $^{16}$O &  1.40E-01  & 4.27E-01  & 3.76E-01  & 1.51E-01  & 4.62E-01  & 5.21E-01\\
 $^{17}$O  & 3.59E-08  & 4.51E-08  & 7.61E-08  & 1.46E-08  & 7.60E-08  & 3.96E-07\\
 $^{18}$O  & 1.13E-09  & 1.80E-09  & 2.42E-09  & 5.24E-10  & 2.04E-09  & 2.57E-08\\
 $^{19}$F  & 7.19E-10  & 1.04E-09  & 1.54E-09  & 2.70E-10  & 5.40E-10  & 3.56E-10\\
 $^{20}$Ne & 1.69E-03  & 6.13E-03  & 1.10E-02  & 4.05E-03  & 3.06E-03  & 2.65E-03\\
 $^{21}$Ne & 1.06E-05  & 1.66E-05  & 2.36E-05  & 4.35E-06  & 7.40E-06  & 4.09E-06\\
 $^{22}$Ne & 2.57E-03  & 1.90E-02  & 2.82E-04  & 7.62E-05  & 2.18E-02  & 2.55E-02\\
 $^{23}$Na & 6.59E-05  & 1.59E-04  & 2.07E-04  & 5.87E-05  & 7.09E-05  & 6.15E-05\\
 $^{24}$Mg & 1.62E-02  & 9.55E-03  & 5.84E-02  & 2.14E-02  & 4.61E-03  & 2.14E-03\\
 $^{25}$Mg & 6.12E-05  & 1.72E-04  & 2.95E-04  & 8.90E-05  & 9.50E-05  & 6.84E-05\\
 $^{26}$Mg & 3.79E-05  & 1.93E-04  & 3.49E-04  & 1.13E-04  & 1.12E-04  & 5.94E-05\\
 $^{27}$Al & 1.00E-03  & 7.23E-04  & 3.22E-03  & 1.17E-03  & 3.45E-04  & 1.99E-04\\
 $^{28}$Si & 1.64E-01  & 5.02E-02  & 3.27E-01  & 2.83E-01  & 3.31E-02  & 1.34E-02\\
 $^{29}$Si  & 9.72E-04  & 6.52E-04  & 3.64E-03  & 1.47E-03  & 3.49E-04  & 1.94E-04\\
 $^{30}$Si  & 2.04E-03  & 9.88E-04  & 7.45E-03  & 2.83E-03  & 4.88E-04  & 1.87E-04\\
 $^{31}$P   & 3.86E-04  & 1.91E-04  & 1.36E-03  & 5.91E-04  & 1.34E-04  & 8.34E-05\\
 $^{32}$S   & 7.74E-02  & 1.87E-02  & 1.07E-01  & 1.21E-01  & 1.27E-02  & 5.76E-03\\
 $^{33}$S   & 3.62E-04  & 1.01E-04  & 1.04E-03  & 5.18E-04  & 7.71E-05  & 6.36E-05\\
 $^{34}$S   & 1.96E-03  & 9.80E-04  & 1.11E-02  & 5.65E-03  & 5.50E-04  & 2.23E-04\\
 $^{36}$S   & 4.10E-07  & 1.49E-07  & 1.30E-06  & 4.78E-07  & 6.19E-08  & 1.97E-08\\
 $^{35}$Cl  & 1.10E-04  & 3.41E-05  & 2.90E-04  & 1.70E-04  & 4.96E-05  & 5.87E-05\\
 $^{37}$Cl  & 2.81E-05  & 7.45E-06  & 7.87E-05  & 5.03E-05  & 8.54E-06  & 1.96E-05\\
 $^{36}$Ar  & 1.21E-02  & 2.54E-03  & 1.21E-02  & 1.93E-02  & 2.19E-03  & 1.49E-03\\
 $^{38}$Ar  & 9.46E-04  & 3.98E-04  & 4.02E-03  & 2.67E-03  & 2.15E-04  & 1.20E-04\\
 $^{40}$Ar   & 6.10E-09  & 1.32E-09  & 1.46E-08  & 5.49E-09  & 7.76E-10  & 8.48E-10\\
 $^{39}$K   & 6.01E-05  & 1.71E-05  & 1.60E-04  & 1.22E-04  & 3.07E-05  & 6.43E-05\\
 $^{40}$K   & 4.34E-08  & 7.17E-09  & 7.75E-08  & 3.88E-08  & 1.71E-08  & 1.95E-07\\
 $^{41}$K   & 4.36E-06  & 1.10E-06  & 1.18E-05  & 8.74E-06  & 1.79E-06  & 1.66E-05\\
 $^{40}$Ca  & 9.73E-03  & 2.04E-03  & 6.82E-03  & 1.69E-02  & 1.71E-03  & 1.52E-03\\
 $^{42}$Ca  & 2.38E-05  & 8.38E-06  & 8.71E-05  & 6.15E-05  & 5.73E-06  & 7.91E-06\\
 $^{43}$Ca  & 7.27E-08  & 1.84E-08  & 1.31E-07  & 1.61E-07  & 5.11E-08  & 3.38E-06\\
 $^{44}$Ca  & 8.40E-06  & 1.94E-06  & 3.31E-06  & 1.59E-05  & 5.66E-06  & 6.22E-05\\
 $^{46}$Ca  & 3.65E-11  & 2.45E-12  & 5.80E-11  & 2.30E-11  & 1.80E-12  & 4.32E-10\\
 $^{48}$Ca  & 2.82E-13  & 1.94E-17  & 4.27E-16  & 1.75E-16  & 2.44E-16  & 3.70E-12\\
 $^{45}$Sc  & 1.33E-07  & 2.55E-08  & 2.38E-07  & 2.29E-07  & 1.59E-07  & 1.59E-05\\
 $^{46}$Ti  & 1.16E-05  & 3.57E-06  & 3.34E-05  & 2.52E-05  & 3.32E-06  & 5.78E-05\\
 $^{47}$Ti  & 5.15E-07  & 1.36E-07  & 7.98E-07  & 1.08E-06  & 8.80E-07  & 8.15E-05\\
 $^{48}$Ti  & 3.65E-04  & 9.23E-05  & 1.65E-04  & 8.16E-04  & 1.38E-04  & 1.34E-03\\
 $^{49}$Ti  & 1.48E-05  & 3.68E-06  & 8.52E-06  & 2.95E-05  & 1.12E-05  & 4.97E-04\\
 $^{50}$Ti  & 1.85E-06  & 9.98E-10  & 3.77E-09  & 1.34E-08  & 2.85E-08  & 7.73E-07\\
 $^{50}$V  & 3.50E-09  & 3.82E-09  & 7.13E-09  & 4.88E-09  & 6.24E-08  & 4.99E-06\\
 $^{51}$V  & 4.04E-05  & 2.15E-05  & 2.95E-05  & 6.52E-05  & 1.28E-04  & 3.01E-03\\
 $^{50}$Cr  & 2.61E-04  & 1.56E-04  & 3.15E-04  & 3.84E-04  & 2.03E-04  & 2.42E-03\\
 $^{52}$Cr  & 4.32E-03  & 2.70E-03  & 2.60E-03  & 8.91E-03  & 2.92E-03  & 9.14E-03\\
 $^{53}$Cr  & 6.60E-04  & 7.12E-04  & 6.03E-04  & 1.08E-03  & 1.28E-03  & 9.12E-03\\
 $^{54}$Cr  & 3.48E-05  & 6.85E-08  & 1.82E-07  & 6.68E-07  & 1.17E-06  & 1.07E-05\\
 $^{55}$Mn  & 6.72E-03  & 7.35E-03  & 5.95E-03  & 7.34E-03  & 1.83E-02  & 4.42E-02\\
 $^{54}$Fe  & 7.77E-02  & 8.15E-02  & 7.01E-02  & 7.40E-02  & 5.98E-02  & 6.55E-02\\
 $^{56}$Fe  & 6.57E-01  & 2.95E-01  & 2.78E-01  & 5.59E-01  & 2.57E-01  & 1.11E-01\\
 $^{57}$Fe  & 2.56E-02  & 1.26E-02  & 1.23E-02  & 1.80E-02  & 1.93E-02  & 2.91E-02\\
 $^{58}$Fe  & 2.29E-04  & 1.36E-05  & 1.27E-05  & 1.57E-05  & 6.42E-05  & 3.63E-04\\
 $^{59}$Co  & 8.29E-04  & 8.20E-04  & 6.70E-04  & 6.36E-04  & 1.31E-03  & 3.22E-03\\
 $^{58}$Ni  & 1.06E-01  & 7.24E-02  & 6.63E-02  & 7.03E-02  & 6.19E-02  & 3.90E-02\\
 $^{60}$Ni  & 9.75E-03  & 1.11E-02  & 8.05E-03  & 8.35E-03  & 6.77E-03  & 3.68E-03\\
 $^{61}$Ni  & 2.73E-04  & 7.20E-05  & 6.95E-05  & 2.05E-04  & 4.92E-05  & 2.30E-04\\
 $^{62}$Ni  & 2.52E-03  & 6.35E-04  & 5.85E-04  & 1.61E-03  & 7.41E-05  & 1.07E-04\\
 $^{64}$Ni  & 2.20E-07  & 6.32E-11  & 1.52E-10  & 1.32E-09  & 7.15E-09  & 3.61E-07\\
 $^{63}$Cu  & 1.77E-06  & 6.49E-07  & 5.99E-07  & 1.34E-06  & 1.36E-06  & 1.21E-05\\
 $^{65}$Cu  & 7.78E-07  & 1.96E-07  & 1.82E-07  & 7.85E-07  & 1.94E-08  & 5.56E-07\\
 $^{64}$Zn  & 1.43E-05  & 3.74E-06  & 3.23E-06  & 1.38E-05  & 9.80E-07  & 5.11E-06\\
 $^{66}$Zn  & 2.37E-05  & 6.39E-06  & 5.58E-06  & 2.13E-05  & 1.02E-07  & 3.36E-07\\
 $^{67}$Zn  & 4.94E-11  & 2.29E-11  & 2.28E-11  & 8.14E-11  & 2.23E-10  & 2.32E-08\\
 $^{68}$Zn  & 9.59E-09  & 2.71E-09  & 2.44E-09  & 9.55E-09  & 2.36E-10  & 6.69E-09\\
 $^{70}$Zn  & 2.36E-14  & 2.60E-19  & 2.89E-18  & 5.53E-17  & 5.78E-15  & 2.93E-11\\
\enddata
\end{deluxetable*}

\section{Discussion: Implications for Chemical Evolution}

In this paper, we present our first results for nucleosynthesis 
in 2D delayed detonation models. The models investigated in this paper represent 
extreme cases\footnote{Note that the O-DDT model is still less 
extreme than the GCD model (Jordan et al. 2008; Meakin et al. 2009) 
in the distribution of the initial bubbles.}; 
the deflagration is initiated either 
at the center (C-DDT) or by off-center bubbles confined in 
the narrow angle with respect to the $z$-axis (O-DDT).  
The important variation, we have not examined in this paper, 
is the deflagration bubbles distributed more or less spherically, 
but at a distance from the center (e.g., Kasen et al. 2009). 
In this case, we expect that the 
propagation of the burning will produce an ejecta structure that falls
in between the O-DDT and the C-DDT models. Some
material in the central region is left unexpanded 
in the deflagration stage, and the detonation can burn the material 
there to the Fe-peak elements. Thus, we expect that nucleosynthesis 
features are similar to the O-DDT model, except for the 
strong angle-dependence seen in the O-DDT model. Depending on the
initial placement of the deflagration bubbles, the deflagration ashes
may stay more confined to the center than in the O-DDT model.

Also, we should note that the treatment of the DDT 
in the hydrodynamic calculation is still preliminary (\S 2.1). 
According to simulations with several different prescriptions, 
especially with different values for $\rho_{\rm DDT}$, 
we believe that the prescription used in this paper results in a relatively 
weak detonation (see also Gamezo et al. 2005; Bravo \& Garc\'ia-Senz 2008). 
Therefore, although the models presented here are
extreme with respect to the ejecta structure, they do not necessarily 
reflect the range of $^{56}$Ni production possible within the framework 
of the delayed detonation model.
The results of an extended survey of delayed detonation models will be
presented elsewhere (F. R\"opke et al.,\ in prep).  
Despite these caveats, the general features found in this paper 
are not expected to be sensitive to such details. 

\begin{deluxetable}{ccccc}
 \tabletypesize{\scriptsize}
 \tablecaption{Masses of radioactive species (\Msun) 
before radioactive decays. 
 \label{tab:tab3}}
 \tablewidth{0pt}
 \tablehead{
   \colhead{Isotopes}
 & \colhead{W7}
 & \colhead{C-DEF}
 & \colhead{C-DDT}
 & \colhead{O-DDT}
}
\startdata
$^{22}$Na & 2.01E-08 & 1.01E-07 & 1.46E-07 & 5.40E-08\\
$^{26}$Al & 5.18E-07 & 1.69E-06 & 2.47E-06 & 8.77E-07\\
$^{36}$Cl & 2.08E-06 & 4.74E-07 & 5.22E-06 & 2.06E-06\\
$^{39}$Ar & 6.79E-09 & 1.53E-09 & 1.69E-08 & 7.75E-09\\
$^{40}$K  & 4.34E-08 & 7.17E-09 & 7.75E-08 & 3.90E-08\\
$^{41}$Ca & 4.35E-06 & 1.10E-06 & 1.18E-05 & 8.85E-06\\
$^{44}$Ti & 8.37E-06 & 1.93E-06 & 3.21E-06 & 1.59E-05\\
$^{48}$V  & 4.32E-08 & 1.68E-08 & 9.76E-08 & 1.09E-07\\
$^{49}$V  & 1.05E-07 & 1.00E-07 & 3.07E-07 & 2.69E-07\\
$^{53}$Mn & 1.64E-04 & 4.93E-04 & 3.38E-04 & 2.25E-04\\
$^{55}$Fe & 1.79E-03 & 4.17E-03 & 2.89E-03 & 1.93E-03\\
$^{60}$Fe & 3.33E-09 & 9.86E-15 & 2.29E-13 & 6.93E-12\\
$^{55}$Co & 4.89E-03 & 3.18E-03 & 3.07E-03 & 5.40E-03\\
$^{56}$Co & 1.21E-04 & 1.18E-04 & 1.06E-04 & 1.04E-04\\
$^{57}$Co & 9.52E-04 & 1.94E-03 & 1.40E-03 & 9.37E-04\\
$^{60}$Co & 4.32E-08 & 5.30E-10 & 1.19E-09 & 3.30E-09\\
$^{56}$Ni & 6.40E-01 & 2.45E-01 & 2.46E-01 & 5.40E-01\\
$^{57}$Ni & 2.46E-02 & 1.06E-02 & 1.09E-02 & 1.71E-02\\
$^{59}$Ni & 4.66E-04 & 7.24E-04 & 5.78E-04 & 4.22E-04\\
$^{63}$Ni & 4.82E-08 & 4.28E-11 & 1.85E-10 & 1.22E-09\\
\enddata
\end{deluxetable}

The nucleosynthesis features in the W7 model have been shown 
to be roughly consistent with the Galactic chemical evolution 
(e.g., Iwamoto et al.\ 1999; Goswami \& Prantzos\ 2000). 
One problem is that the ratio 
$^{58}$Ni/$^{56}$Fe is too large in the W7 
to compensate the over-solar production of $^{58}$Ni in core-collapse SNe. 
The problem is basically solved by introducing a delayed
detonation. Affecting primarily lower density material, it 
produces a large amount of $^{56}$Ni but virtually no $^{58}$Ni due to
the inefficiency of electron captures here. We see that the ratio 
$^{58}$Ni/$^{56}$Fe is slightly decreased in the O-DDT 
model compared to the W7, marginally satisfying 
the constraint from the chemical evolution 
[($^{58}$Ni/$^{56}$Fe)/($^{58}$Ni/$^{56}$Fe)$_{\odot}$ $\sim 3$]. 
The situation should become better if the detonation burns more
material to NSE than in the present prescription. 
We should also note one limitation in the present analysis, i.e., 
the treatment of the temperature in the NSE regime. 
The temperature is extracted assuming $Y_{\rm e} = 0.5$, 
which should introduce some errors when the electron capture 
reactions are active. This could to some extent affect the 
nucleosynthesis in the initial deflagration phase (\S 3.4), 
and therefore the ratio $^{58}$Ni/$^{56}$Fe. 

The intermediate mass elements can also be used to 
constrain SN Ia models. Assuming a typical fraction of 
$\sim 20$\% for the frequency of SNe Ia as compared to 
core-collapse SNe, then 
(Si/Fe)/(Si/Fe)$_{\odot}$ $\lsim 0.5$ is required to 
compensate the over-solar production in core-collapse SNe. 
The O-DDT model has the ratio (Si/Fe)/(Si/Fe)$_{\odot} \sim 1$; 
thus, the model can explain at most a half of SNe Ia. 
The C-DDT is not at all favored, since (Si/Fe)/(Si/Fe)$_{\odot} \sim 2$, 
i.e., it produces too much IMEs relative to Fe. 
The deflagration models (W7 and C-DEF) do not have the problem 
in the Si/Fe ratio [(Si/Fe)/(Si/Fe)$_{]\odot} < 0.5$].  
In the O-DDT model, all the elements heavier than N are consistent 
with W7 within a factor of two.

In reality, the solar abundances are the superposition 
of contributions from SNe at various metallicities, and 
we expect that the "average" SNe Ia to occur at a 
metallicity less than solar. We indeed expect some improvement if 
we consider a lower metallicity (see e.g., Travaglio et al.\ 2005): 
a decreased metallicity should result in a smaller  
$^{58}$Ni/$^{56}$Fe ratio in the delayed detonation models. 
Also, the smaller metallicity is expected to lead to 
the smaller amount of $^{54}$Fe. 
Mg and Al are also affected: If the metallicity is 10\% of the solar 
value, then the amount of $^{24}$Mg could increase by $\sim 50$\%, 
and that of $^{27}$Al could decrease by a factor of a few. 
These changes do not conflict with the Galactic chemical evolution, 
since the ratios (Mg, Al)/Fe are much smaller than the solar values 
in the O-DDT models. 

Summarizing, the hypothesis that 
about half of SNe Ia are represented by the extreme O-DDT model, 
is not rejected by the chemical evolution arguments. Note that 
this should also apply to globally symmetric, but off-center, delayed
detonation models  
to some extent, as these models should share the basic feature that the 
central region is burned to Fe-peak and $^{56}$Ni by the detonation 
(\S 4). Therefore, multi-D delayed detonation models can potentially
account for a majority of  
SNe Ia.

\section{Concluding Remarks}

We have presented results of 
the detailed nucleosynthesis calculations for 2D delayed detonation models, 
focusing on an extremely off-center model (O-DDT model). 
Features are different from classical 1D models. 
Unlike a globally symmetric delayed detonation model following the 
central ignition of the deflagration flame (C-DDT model), 
the detonation wave proceeds both in the high density region near the 
center of a white dwarf and in the low density region near the surface. 
Thus, the resulting yield is a mixture of different explosive 
burning products, from carbon-burning at low densities to 
complete silicon-burning at the highest densities, as well as 
the electron-capture products from the deflagration stage. 

The evolution of the deflagration flame can be different in 2D and in 3D 
simulations (e.g., R\"opke et al. 2007b). We believe that the global feature found in this 
study, i.e., the detonation wave propagating into the innermost region, would not be changed 
substantially in 3D simulations. 
However, some details would be affected: For example, 3D simulations tend to result 
in stronger deflagration than in 2D simulations for similar initial conditions 
(e.g., Travaglio et al. 2004a). As the stronger deflagration 
is followed by the weaker detonation in our DDT models, this would
result in a shift 
of the overall energetic and the nucleosynthesis production as compared to the 2D models. 
On the other hand, 
although in our 2D simulations the detonation 
wave has to burn around the deflagration ash blob before reaching to the center, the deflagration 
ash blob may have holes, depending on the ignition geometry 
through which the detonation wave can directly penetrate into the center in 3D simulations. 
Thus, future 3D simulations are important  
to obtain the robust model predictions as a function of the model input (e.g., 
distribution of the deflagration ignition sparks). 

The yield of the O-DDT model largely satisfies constraints from the Galactic chemical 
evolution despite the low DDT density as compared to 
1D delayed detonation models. This is a result of qualitatively different behavior of the detonation 
propagation in 1D and multi-D models; the outward propagation in former 
(which is also the case in the C-DDT model), 
while the inward propagation in the latter. 

The O-DDT model could thus account for 
a fraction of SNe Ia (especially bright ones). As less-extreme 
(more spherical) models are expected to share the common properties in 
the integrated yield, the multi-D delayed detonation model could potentially account for a 
main population of SNe Ia. 

The delayed detonation models also provide a characteristic layered structure, unlike 
multi-dimensional, especially 2D, deflagration models
\footnote{Note that the 3D deflagration models can also potentially produce 
the layered structure (R\"opke et al. 2007b).}. 
In the O-DDT model, 
the region filled by electron capture 
species (e.g., $^{58}$Ni, $^{54}$Fe) shows a large off-set, 
and the region is within a shell above 
the bulk of the $^{56}$Ni distribution near the center. 
These can be directly tested by observations. 

We note that the distribution of the nucleosynthsis products 
in our 2D DDT models is somewhat different from the result of 
Bravo \& Garc\'ia-Senz (2008). Their 3D DDT models lack a clear 
abundance stratification, and they are characterized by 
large mass fractions of Fe-peak elements near the surface regions. 
In contrast, the 3D DDT models of R\"opke \& Niemeyer (2007c), do produce 
a layered structure with the surface dominated by IMEs -- similar to our 
present 2D models (\S 3.5). 
It seems that (1) the deflagration is stronger in Bravo \& Garc\'ia-Sent 
(2008) than in our 2D models and in 3D models of R\"opke \& Niemeyer (2007c), 
and (2) the detonation wave is mainly propagating inward in Bravo \& Garc\'ia-Senz (2008) 
although it is propagating both inward (producing Fe-peak elements) and outward (producing IME) 
in our simulations. As a result, the amount of unburned material is smaller in our models. 
The cause of the different flame propagation is not clear, but likely due to different 
treatment of thermonuclear flames. The overall abundance distribution of the 3D DDT models 
of Gamezo et al. (2005), on the other hand, is similar to our 2D models, producing the 
stratified configuration. In their simulations, they initiated the detonation from the center 
at relatively high DDT density, and the detonation wave propagates outward, producing the 
layered composition structure as the temperature drops following the detonation propagation. 
This is, indeed, quite different from our models, in which the detonation is initiated at relatively low 
DDT density but propagates inward to the high density central region. 

The observational consequences from similar models have been discussed by Kasen et al. 
(2009) for the early photospheric phase 
(see also Hillebrandt et al.\ 2007; Sim et al.\ 2007). 
Here, we summarize some additional, expected 
observational characteristics, especially in the late-time nebular phase (taken after a few hundred days). We emphasize that the late-time spectroscopy 
is currently the most effective way to hunt for the signature of 
the DDT model in the innermost region. See also Maeda et al. (2010) 
who discussed the following points in details. 
\begin{itemize}
\item {\bf Abundance Stratification and carbon near the surface: }
Overall, the stratified composition structure obtained in our 2D models  
is consistent with the result of 
the "abundance tomography" (Stehle et al. 2005; Mazzali et al. 2008). 
Spectrum synthesis models as compared to the observed early-phase 
photospheric spectra show that the mass fraction of carbon at 
$\sim 10,000 - 14,000$ km s$^{-1}$ 
should be smaller than $0.01$ (e.g., Branch et al. 2003; Thomas et al. 2007; 
Tanaka et al. 2008). This constraint 
is satisfied by our DDT models at $10,000$ km s$^{-1}$, although 
it is marginal at $14,000$ km s$^{-1}$. The latter could, however, be improved 
by changing the DDT criterion such that the transition on 
average proceeds at higher densities (e.g., Iwamoto et al. 1999). 
Another interesting observational target is the inhomogeneous 
distribution of the unburned C+O pockets near the surface both in C-DDT and 
O-DDT models (\S 3.5), although the total amount of such unburned material 
is small in the DDT models. Detectable carbon absorption lines may appear 
only when the observed line-of-sight intersects a large number of such unburned 
pockets, which may explain the low frequency of  SNe Ia showing the C absorption lines 
detected (e.g., Tanaka et al. 2006). 
\item {\bf [O~I]~$\lambda\lambda$6300,6363: }
The main problem in the pure deflagration model is existence of unburned 
carbon and oxygen mixed down to the central region. This should
produce a strong [O~I]~$\lambda\lambda$6300, 6363 doublet in late-time spectra, although 
no such signature has been detected in the observations (Kozma et al.\ 2005). 
This tension is alleviated in 3D deflagration models, with  
the mass fraction of unburned elements going down to 10 \% 
(R\"opke et al.\ 2007b) or even to smaller values 
(e.g., Travaglio et al.\ 2004a). 
The DDT model does not have this problem, as unburned material
 near the center is burned in the subsequent detonation. 
A detailed study of late-time nebular spectra 
should provide us with information on this issue. 
\item {\bf Line shifts: }
Profiles of nebular emission lines 
can be used to effectively trace the distribution of the burning products, 
as is proven to be efficient for core-collapse SNe (Maeda et al.\ 2002; 
Mazzali et al.\ 2005; Maeda et al.\ 2008; Modjaz et al.\ 2008; Taubenberger et al.\ 2009). 
The O-DDT model predicts that (1) the distribution of stable Fe-peak elements is off-set 
following the asymmetric deflagration flame propagation, while 
the detonation products (e.g., a large fraction of $^{56}$Ni) 
are distributed more or less in a spherical manner (Figs.~5 and 12). 
Recently, Maeda et al. (2010) found that the expected variation of 
the line wavelength is seen in nebular spectra of SNe Ia, indicating that 
the above configuration can be relatively common in SNe Ia. 
\item {\bf Line profiles: }
Detailed profiles of the nebular emission lines can be used to infer the distribution of 
emitting ions. It has been suggested to use Near-Infrared (NIR) 
spectroscopy to hunt for the geometry, 
as NIR [Fe~II] lines are not severely blended (e.g., H\"oflich et al.\ 2004; Motohara et al.\ 2006). 
This can also be done using several IR lines like [Co III] 11.88$\micron$ (Gerardy et al. 2007). 
The best data to date have been 
obtained for SN 2003hv, showing a hole in the distribution of 
$^{56}$Ni\footnote{Note, however, that the central wavelengths of the emission lines are 
shifted with respective to the explosion center, and this "line shift" can be well explained by 
the off-set DDT scenario (Maeda et al. 2010).}. Because the mixing is introduced by the 
deflagration, such a hole is difficult to understand in the present models 
(and most of SN Ia explosion models; but see Meakin et al. 2009). This issue remains unresolved, 
and need further study preferentially in 3D simulation. 
\end{itemize}

\acknowledgements
We would like to thank Ken'ichi Nomoto for kindly providing the thermal history of 
the W7 model. This research has been supported by World Premier International Research 
Center Initiative (WPI Initiative), MEXT, Japan. 
The work of K.M.\ is also supported through 
the Grant-in-Aid for Young Scientists (20840007) of Japanese Society for 
Promotion of Science (JSPS). 
The work of F.K.R.\ is supported 
through the Emmy Noether Program of the German Research Foundation (DFG;
RO~3676/1-1) and by the Cluster of Excellence ``Origin and
  Structure of the Universe'' (EXC~153).
The work by F.-K.T.\ is supported by the Swiss National Science 
Foundation(SNF) and the Alexander von Humholdt Foundation.
The calculations have been performed on 
IBM Power5 system at Rechenzentrum Garching (RZG) 
of the Max-Planck Society.

\end{document}